%% file: main.tex
\documentclass{optica-article}

\journal{opticajournal} 

\articletype{Research Article}

\usepackage{lineno}
\usepackage{tikz}

\usepackage{mathtools}
\usepackage{graphicx}
\usepackage{comment}
\DeclareMathOperator{\sinc}{sinc}
\DeclareMathOperator{\rect}{rect}

\DeclareMathOperator{\ceil}{ceil}

\begin{document}

\title{Attenuation scaling and error analysis of 2f and 4f architectures for free-space optical matrix-vector multiplication}

\author{Dawson Lyles,\authormark{1} Spencer LaVere Smith,\authormark{2,*} }

\address{\authormark{1}Department of Physics, University of California Santa Barbara, Broida Hall, Santa Barbara,
California 93106-9530, USA\\
\authormark{2}Department of Electrical and Computer Engineering, University of California Santa Barbara, 2002 BioEngineering Building, Santa Barbara,
California 93106-5100, USA\\}

\email{\authormark{*}sls@ucsb.edu} 


\begin{abstract*} 
Free-space optical computing has been suggested as a scalable, high speed, and energy efficient platform for performing matrix-vector multiplication (MVM). We present two free-space optical approaches for MVM, called the $2f$ and $4f$ architectures, and model them using wave optics simulations. After constraining the optical modulator in our models to have a maximum gain limit, we use our simulations to compare $2f$ and $4f$ MVM performance in terms of computational error and optical signal attenuation per MVM. We examine how $2f$ and $4f$ signal attenuation per MVM scales with increasing MVM problem size for different statistical distributions of matrix elements and compare to the expected attenuation from a universal multiport interferometer (UMI), commonly used in integrated photonics for MVM. We find that the $2f$ and $4f$ architectures scale more favorably to large problem sizes, experiencing many orders of magnitude less attenuation than UMIs for matrix dimension above a thousand elements. We furthermore examine how varying modulator space-bandwidth product and output slit aperture affect $2f$ and $4f$ attenuation and computational error across different distributions of matrix elements. We conclude that the preference of $2f$ or $4f$ MVM depends on the statistics of the matrix used, but that $4f$ may provide more flexibility than $2f$. 
\end{abstract*}

\section{Introduction}\label{sec:intro}
Photonic computing has been identified as a high speed, efficient, and scalable way to accelerate large-scale linear transformations necessary for AI and other data-intensive algorithms \cite{WU_OC_for_AI,fu_ONN_prog_and_chall}. The many degrees of freedom of light (e.g. polarization, spatial and temporal frequencies, phase, and amplitude), paired with weak optical nonlinearity in most materials, provides extremely high parallelism for communication and linear data processing \cite{mcmahon_physics_2023}. In addition to the advantage of massive parallelism, optical computing systems can operate at extremely low power \cite{TW_less_than_1_photon, kim_low_power,homodyne_free_space}. Since the 1970s \cite{Heinz}, various photonic architectures have arisen specialized to matrix-vector multiplication (MVM), the core operation in linear computing. Today, photonic MVM platforms span free-space \cite{Spall:20,arb_linear_trans,homodyne_free_space} and integrated photonic architectures using microring resonator arrays \cite{microring_MVM} and universal multiport interferometer (UMI) meshes \cite{UMI_MVM}. For all of these approaches, an array of optical modes, encoding an input vector, is transformed by spatially-distributed modulator components encoding the matrix weights. 

We limit our focus to MVM architectures that transform a set of input \textit{spatial} optical modes, encoding an input vector, into a set of output \textit{spatial} modes, encoding an output vector. The complex amplitude of each spatial mode encodes the numerical value of each vector element. Among these approaches, UMIs \cite{reck_zeilinger,Clements:16} rely on repeated mixing and modulation of spatial modes across many interferometric layers to perform MVM. There also exist free-space MVM approaches that transform spatial modes using a single modulating layer. In fact, it has been shown that any arbitrary complex-valued linear transformation can be approximated using a single modulating layer, as long as the modulating layer has enough pixels \cite{arb_linear_trans}. In this paper, we examine two such "single-layer" architectures for MVM originally suggested by Tamura et al. \cite{Wyant} and Heinz et al. \cite{Heinz}, which we call the $2f$ and $4f$ architectures, respectively. The $2f$ and $4f$ architectures have been used to build optical discrete Fourier transform systems in \cite{Goodman_DFT} and \cite{ZHANG_4f_DFT}, respectively, and optical neural networks in \cite{Zuo_2f_ONN_neuron,TW_less_than_1_photon,homodyne_free_space} and \cite{Chang_4f_CNN, bernstein2023single,Yan_4f_CNN}, respectively. While our definition of the $2f$ architecture exactly matches that of \cite{Wyant,Spall:20}, we slightly modify the method of \cite{Heinz} so that the modulator in our $4f$ setup can apply a complex-valued transmittance mask (enabling phase and intensity control of light). 

Suppose we have an array of spatial modes encoding an input vector which is transformed into an array of output spatial modes after passing through an optical MVM system. If the modulator components within the system cannot amplify the intensity of the spatial modes (e.g. the modulator components are passive), then the energy of the input optical signal is expected to be attenuated during MVM due to propagation and insertion losses of the modulating components. Therefore, the depth (number of modulating layers) of an MVM system should greatly impact the expected loss of a signal from input to output. We will show that the expected loss of a multilayer UMI scales drastically faster than single-layer $2f$ and $4f$ architectures. 

In this paper, we use wave optics simulations to compare the optical signal attenuation per MVM and expected computational error of the $2f$ and $4f$ MVM architectures. We examine how $2f$ and $4f$ attenuation scales with increasing matrix dimension (input vector length) and compare to the attenuation scaling of UMIs. Our results show that $2f$ and $4f$ architectures are far more scalable to large MVMs and more amenable to cascaded MVM than UMIs. We also show that $2f$ and $4f$ attenuation scalings depend on the statistics of the transforming matrix elements. Furthermore, we demonstrate that the $4f$ architecture may be more flexible than the $2f$ architecture, allowing computational error to be reduced without increasing attenuation. However, we cannot definitively conclude whether the $2f$ or $4f$ architecture is preferable for MVM, as the preference for one or the other depends on the statistics of the matrix and the amount of error and signal attenuation we can tolerate in our computations. Our findings suggest that single-layer MVM approaches, such as $2f$ and $4f$, are more scalable than multilayer approaches such as UMIs.

\begin{figure*}[h]
  \noindent
  \centering \makebox[\textwidth][c]{ \includegraphics[width=1.2\textwidth]{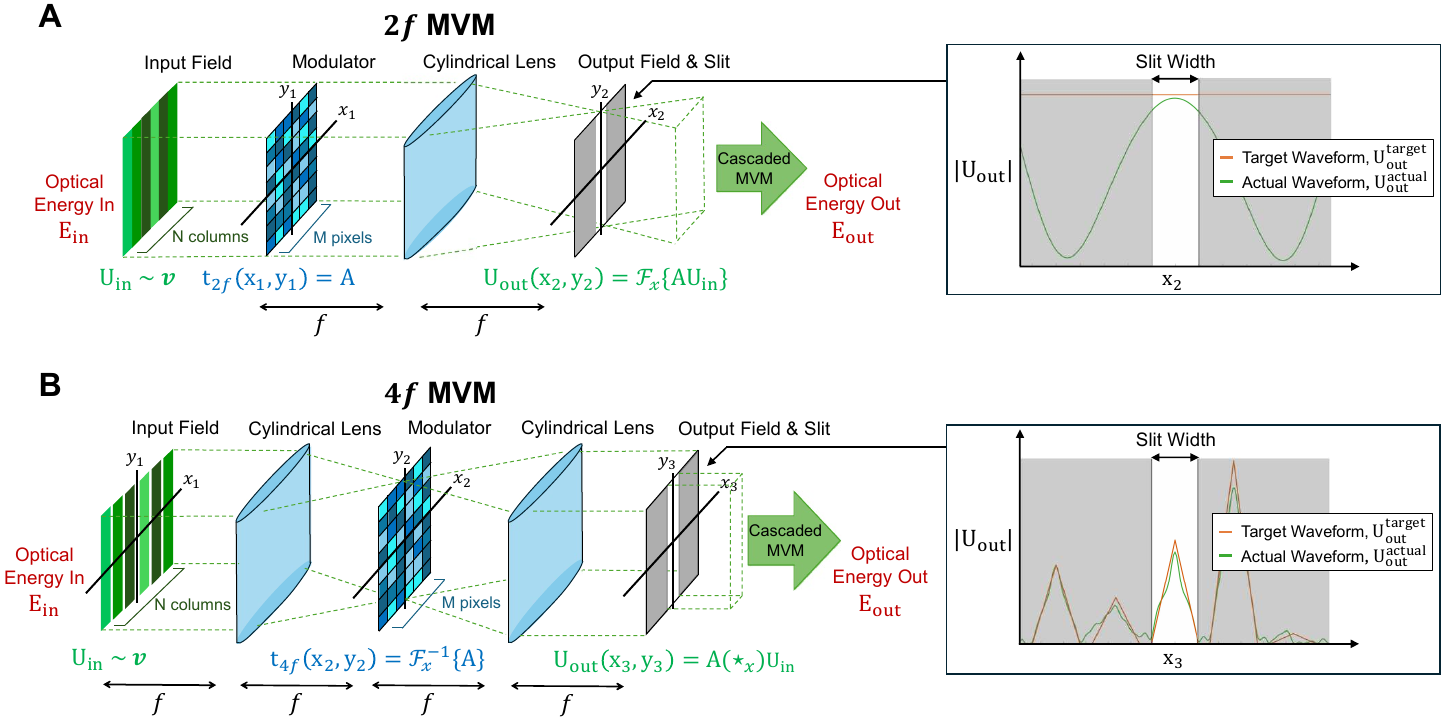}}
  \caption{\label{fig:2f_4f_mvm} (A) Schematic of a $2f$ MVM system. An input optical scalar field $U_{\mathrm{in}}$ in the input plane has magnitude and phase profiles which together encode an input vector, $\boldsymbol{v}$. Each column of the input field $U_{\mathrm{in}}$ encodes an element of the input vector, shown as stripes of varying shades of green, creating a copy of $\boldsymbol{v}$ for each row of a matrix $\boldsymbol{A}$. The field encounters a modulator which applies the transmittance function $t_{2f}(x,y)$. The modulator pixels encode the values of the matrix, $\boldsymbol{A}$. A cylindrical lens one focal length $f$ behind the modulator performs a Fourier transform, along the $x$-dimension, of the modulated field onto its back focal plane to give the output field $U_{\mathrm{out}}$ in its back focal plane. A slit selects out the central column of $U_{\mathrm{out}}$, which encodes the MVM solution, $\boldsymbol{A} \boldsymbol{v}$, and the field can continue to be multiplied by another matrix in cascaded MVM. Boxed on the right is a plot of a horizontal $x$-cross section of $|U_{\mathrm{out}}|$ in the slit plane, where the field not selected by the slit is grayed out. Traces in the boxed plot encode a single component of the solution $\boldsymbol{A} \boldsymbol{v}$. The desired target waveform (light orange) is compared to the actual received field (green). (B) Schematic of a $4f$ MVM system. An incoming field $U_{\mathrm{in}}$ encodes an input vector, $\boldsymbol{v}$, while the transmittance function of the modulator, $t_{4f}(x,y)$, encodes the Fourier transform of the matrix $\boldsymbol{A}$. The addition of a second cylindrical lens (to give four total focal lengths) means the $4f$ system performs a convolution between each copy of vector $\boldsymbol{v}$ and each row of $\boldsymbol{A}$ to obtain the MVM solution in the output field, $U_{\mathrm{out}}$. The slit selects out the desired column of $U_{\mathrm{out}}$ encoding $\boldsymbol{A} \boldsymbol{v}$. The roughly triangular traces in the boxed plot to the right encode a single component of the solution $\boldsymbol{A} \boldsymbol{v}$. Our target waveform is perfectly triangular (light orange), but the actual received field (green) deviates from this shape.}
\end{figure*}

\section{2f and 4f MVM} \label{sec:2f_and_4f_MVM}
A matrix-vector product $\boldsymbol{A}\boldsymbol{v}$ is performed by calculating the dot product between vector $\boldsymbol{v}$ and each row of $\boldsymbol{A}$. We present two architecturees for free-space optical MVM, called the $2f$ and $4f$ architecturees, which perform dot products between $\boldsymbol{v}$ and all matrix rows in parallel to compute $\boldsymbol{A}\boldsymbol{v}$ in a single pass. We provide overviews of these two architectures, derive expressions for their expected output fields using wave optics theory, and verify these analytic results using simulations. We also remark on the error inherent to $2f$ and $4f$ MVM.

\subsection{2f MVM}
The $2f$ architecture implements the following method for performing the matrix-vector product $\boldsymbol{A}\boldsymbol{v}$, described in detail in Method 1 of the Supplementary Information:
\begin{enumerate}
    \item Vector $\boldsymbol{v}$ is multiplied elementwise by each row of $\boldsymbol{A}$.
    \item A Fourier transform is applied to each row of elementwise products.
    \item The zeroth-order Fourier component for each row is the sum of all elementwise products in that row. The resulting sums for each row are the dot products between $\boldsymbol{v}$ and the rows of $\boldsymbol{A}$. The vector of sums is the solution $\boldsymbol{A}\boldsymbol{v}$.
\end{enumerate}
We use a Fourier transform to sum elements together
because the Fourier transform is unitary and thus can be implemented using linear, passive optical components (e.g. a lens). Using a linear optical device, it is impossible to losslessly couple all the energy of two or more input orthogonal optical modes into a single output mode \cite{miller_mode_converters}. This process also violates the second law of thermodynamics. Therefore, we cannot use a linear optical device to perfectly "add" together the energies or amplitudes of two orthogonal optical modes. Instead, we losslessly transform the input modes to the Fourier basis, where one of the output modes is proportional to the sum of amplitudes of the input modes. 

The $2f$ architecture reflects the setup described in \cite{Wyant,Spall:20}. The $2f$ optical system (see Fig.~\ref{fig:2f_4f_mvm}(A)) consists of a modulator, a cylindrical lens, and a slit. An incoming optical scalar field $U_{\mathrm{in}}(x, y)$ encodes the elements of an $N$-length vector $\boldsymbol{v}$ in its magnitude and phase profiles in the $xy$-\nobreakdash plane transverse to propagation. The magnitude (phase) of the field at any point encodes the magnitude (phase) of the corresponding complex-valued vector element $v_m$. Each vertical column of $U_{\mathrm{in}}$ is assigned a vector element $v_m$, such that a copy of vector $\boldsymbol{v}$ is multiplied with each row of matrix $\boldsymbol{A}$ encoded on the modulator. $U_{\mathrm{in}}$ is analogous to matrix $\boldsymbol{V}$ of Method 1 (see Supplementary Information). Each modulator pixel has a transmission coefficient $t_{mn}$ corresponding to an element of $\boldsymbol{A}$. The resulting transmittance function of the modulator $t_{2f}(x, y)$ alters the magnitude and phase profiles of $U_{\mathrm{in}}$, performing elementwise products between each row of $\boldsymbol{A}$ and each copy of $\boldsymbol{v}$. The cylindrical lens maps the modulated field in its front focal plane to the Fourier transform (along the $x$-dimension) of the modulated field in its back focal plane to find the sum along each row of elementwise products. We call the field in the back focal plane the output field $U_{\mathrm{out}}$. A slit centered at $x_2 = 0$ captures the zeroth-order Fourier component of the modulated field $U_{\mathrm{in}} \, t_{2f}$, selecting out a vertical strip encoding the MVM solution $\boldsymbol{A} \boldsymbol{v}$. The slit plane lies in the Fourier plane of the cylindrical lens, hence a slit of full-width $\mathrm{SW}$ in real space forms a spatial frequency aperture of bandwidth $\Delta \nu_x = \mathrm{SW}/(\lambda f)$ selecting frequencies $\nu_x \in [-\mathrm{SW}/(2 \lambda f)\, , \, \mathrm{SW}/(2 \lambda f)]$.  We call $\Delta \nu_x$ the "slit bandwidth." 

\subsubsection{Analytic Calculations: 2f Architecture} \label{subsec:2f_MVM_analytic}
We calculate the evolution of a horizontal $x$-cross section of the input optical field $U_{\mathrm{in}}$ through the $2f$ system using scalar wave optics theory. We can restrict to one dimension without loss of generality since all focusing occurs along the horizontal $x$\nobreakdash-dimension and further optics can be included to perfectly image along the $y$\nobreakdash-dimension. All fields and transmittance masks will be one-dimensional functions. The following calculations therefore correspond to optically computing the dot product between an $N$-length input vector $\boldsymbol{v}$ and a single $N$-length row vector, $\boldsymbol{a}$, of matrix $\boldsymbol{A}$. We refer to Fig.~\ref{fig:2f_4f_mvm}(A) for all coordinate axes labels and function names. 

The modulator is composed of $M$ rectangular pixels, each of full-width $W$ and spaced at pixel pitch $r$. The transmittance of each pixel is programmed to encode the elements of vector $\boldsymbol{a}$ according to
\begin{equation}\label{eq:modulator_t_2f}
    t_{2f}(x_1) = \sum_{m=1}^{M} a_{k(m)} \rect \! \left(\frac{x_1 - mr - X_0}{W}\right);\quad  k(m) = \ceil \! \left (\frac{mN}{M} \right), 
\end{equation}
where we have included a centering parameter $X_0$. Note that $M \geq N$ in general, so multiple pixels can be encoded with the same element $a_n$ to create superpixels, hence the indexing function $k(m)$. We use all $M$ modulator pixels to encode $\boldsymbol{a}$ and shape the input field $U_{\mathrm{in}}$ so that each of its pulses encoding element $v_n$ is aligned with the corresponding pixels on the modulator encoding $a_n$. 
As such, in the $xy$-\nobreakdash plane transverse to propagation, the input optical field $U_{\mathrm{in}}(x)$ takes the form of a rectangular function motif, $\rect(x)$, of width $W$ repeated $M$ times at regular spacing $r$, with each rectangular pulse scaled appropriately to encode $\boldsymbol{v}$. The optical field in the plane immediately before the modulator is therefore
\begin{equation} \label{eq:U_{in,2f}}
    U_{\mathrm{in}}(x_1) =
    \sum_{m=1}^{M} v_{k(m)} \rect \! \left(\frac{x_1 - mr - X_0}{W}\right).
\end{equation}
The field immediately following the modulator plane is 
\begin{equation}
\begin{aligned}
    U_1(x_1) & = U_{\mathrm{in}}(x_1) \, t_{2f}(x_1) \\ & = \sum_{m=1}^{M} a_{k(m)}v_{k(m)} \rect \! \left(\frac{x_1 - mr - X_0}{W} \right).
\end{aligned}
\end{equation}
Referring to Eq.~(S13) of the Supplementary Information, the output field in the back focal plane of cylindrical lens is the Fourier transform of the field $U_1$
\begin{equation} \label{eq:predicted_U_out_2f}
    \begin{aligned}
        U_{\mathrm{out},2f}(x_2) & = \mathcal{F}_x\{U_1\} = \frac{1}{\sqrt{\lambda f}} \int_{-\infty}^{\infty} \exp \! {\left(-i\frac{2\pi}{\lambda f}x_2 \xi \right)} \, U_1(\xi) \, \mathrm{d}\xi \\
        & = \frac{W}{\sqrt{\lambda f}} \exp \! {\left(-i \frac{2\pi}{\lambda f}X_0 x_2 \right)} \sinc \! \left(\frac{\pi W x_2}{\lambda f} \right) \sum_{m=1}^{M} a_{k(m)} v_{k(m)} \exp \! {\left(-i \frac{2\pi}{\lambda f}m r x_2 \right)},
    \end{aligned}
\end{equation}
where $\sinc(x) = \sin(x)/x$. The field in the output plane consists of a $\sinc$ function envelope multiplied by a modulating function which carries the dot product solution $D = \boldsymbol{a} \cdot \boldsymbol{v}$. The output slit passes the output field within the interval $x_2 \in [-\mathrm{SW}/2  ,   \mathrm{SW}/2]$, where $\mathrm{SW}$ is the slit full-width, forming a frequency aperture that passes spatial frequencies $\nu_x \in [-\mathrm{SW}/{(2 \lambda f)} , \mathrm{SW}/{(2 \lambda f)}]$. We define the "slit bandwidth" $\Delta \nu_x = \mathrm{SW}/(\lambda f)$ as the range of spatial frequencies passed by the slit. 

\subsubsection{2f Dot Product Example}\label{susbec:2f_example}

\begin{figure*}[htbp]
  \noindent 
  \centering {\includegraphics[width=0.7\textwidth]{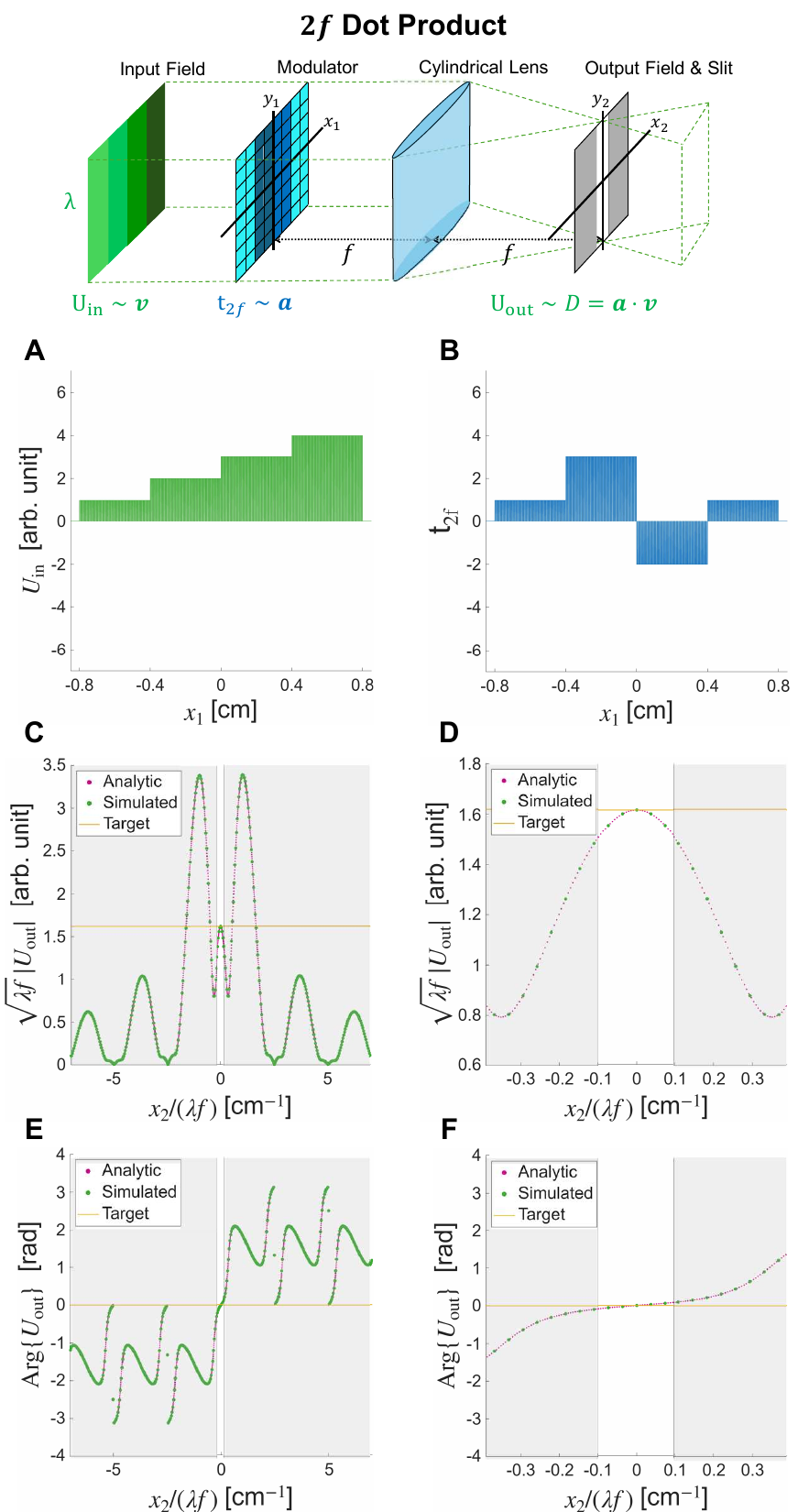}} \caption{\label{fig:2f_dot_prod_fields_example}
  Top: Diagram of $2f$ architecture performing vector dot product $D = \boldsymbol{a} \cdot \boldsymbol{v}$ showing input field $U_{\mathrm{in}}$, modulator transmittance $t_{2f}$, and output field $U_{\mathrm{out}}$ and relevant planes.
  (A) Input optical field $U_{\mathrm{in}}$ immediately before the modulator encoding vector $\boldsymbol{v} = ( 1, 2, 3, 4)^T$. The comb-like appearance of the plot trace matches that of the modulator transmittance. (B) Modulator transmittance function $t_{2f}$ residing in the $x_1y_1$-\nobreakdash plane and encoding vector $\boldsymbol{a} = (1, 3, -2, 1)^T$. The comb-like appearance of the plot trace is due to the $M = 400$ individual pixels of the modulator. (C) Magnitude of the output field $U_{\mathrm{out}, 2f}$ in the slit plane $x_2y_2$ encoding solution $D = \boldsymbol{a} \cdot \boldsymbol{v}$, compared to the desired (but physically impossible) target field (light orange). Numerically simulated results (green) confirm the output field predicted analytically (magenta) from Eq.~(\ref{eq:predicted_U_out_2f}). The portion of the field not passed by the slit is grayed out. (D) Zoomed in version of (C). The slit bandwidth is $\Delta \nu_x = 0.2 \, \mathrm{cm}^{-1}$. (E) Phase of the output field $U_{\mathrm{out}, 2f}$ in the slit plane $x_2y_2$. (F) Zoomed in version of (E). }
  \end{figure*}

As an example, consider the $2f$ architecture performing the dot product of two $N = 4$ length vectors $D=\boldsymbol{a} \cdot \boldsymbol{v} = (1, 3, -2, 1)^T \cdot (1, 2, 3, 4)^T = 5$, where $\boldsymbol{v}$ is encoded in the optical field $U_{\mathrm{in}}$ and $\boldsymbol{a}$ is encoded in transmittance function $t_{2f}$ according to Sec.~\ref{subsec:2f_MVM_analytic} (see Fig.~\ref{fig:2f_dot_prod_fields_example}(A) and (B)). In this example, the modulator is $L_{\mathrm{mod}} = 1.6 \, \mathrm{cm}$ wide and composed of $M = 400$ pixels. The pixel width is $W = Fr = 17(40 \, \mathrm{\mu m})/21 \approx 32.38 \, \mathrm{\mu m}$, where $F=17/21$ is the modulator pixel fill-factor and $r=40 \, \mathrm{\mu m}$ is the pixel pitch. The slit bandwidth is chosen as $\Delta \nu_x = 0.2 \, \mathrm{cm}^{-1}$.

After passing through the \textit{2f} system, the output field in the slit plane $U_{\mathrm{out}, 2f}$ (see Fig.~\ref{fig:2f_dot_prod_fields_example}(C--F)) encodes the dot product $D = \boldsymbol{a} \cdot \boldsymbol{v}$. The numerically simulated $U_{\mathrm{out},2f}$ from the sampled $U_{\mathrm{in}}$ and $t_{2f}$ functions confirms the analytically predicted $U_{\mathrm{out}, 2f}$ from Eq.~(\ref{eq:predicted_U_out_2f}). The value of the field at $x_2=0$ encodes the dot product solution according to Eq.~(\ref{eq:predicted_U_out_2f}): $\sqrt{\lambda f}\, U_{\mathrm{out}, 2f}(x_2=0) = (\boldsymbol{a} \cdot \boldsymbol{v})(MW/N) \approx 1.619 \, \mathrm{cm}$. In this example, because vectors $\boldsymbol{a}$ and $\boldsymbol{v}$ are real-valued, then the field immediately after the modulator is real-valued. Since the lens Fourier transforms this modulated field, the output field in the slit plane $U_{\mathrm{out}, 2f}$ is conjugate symmetric (i.e. $U^*_{\mathrm{out}, 2f}(x_2) = U_{\mathrm{out}, 2f}(-x_2)$). This implies that the dot product solution at $x_2 = 0$ lies at a local extremum of the field magnitude; i.e. the field magnitude is flat locally at $x_2 = 0$ (see Fig.~\ref{fig:2f_dot_prod_fields_example}(D)). On the other hand, the phase may vary rapidly around $x_2 = 0$ depending on the dot product performed. 

The slit in the output plane $x_2y_2$ passes the spatial frequencies $\nu_x \in [-\mathrm{SW}/{(2 \lambda f)} ,  \mathrm{SW}/{(2 \lambda f)}]$. Note that the slit could be replaced with an array of photodiodes to detect the field over the slit width. Although we only desire the zeroth-order Fourier component at $x_2 = 0$, which encodes the dot product solution, the finite width of the slit or detector pixel means unwanted frequencies are also passed/detected. The variation of $U_{\mathrm{out},2f}$ over the slit width introduces error in the optically-computed dot product. Ideally, we would want the field to be flat over the slit width to eliminate error (see "Target" field of Fig.~\ref{fig:2f_dot_prod_fields_example}(C--F)), but this field flatness is physically impossible due to the wave nature of light. In this example, the target field within the slit has the form 
\begin{equation}\label{eq:target_U_out_2f}
U_{\mathrm{out},2f}^{\mathrm{target}}(x_2) = \frac{(\boldsymbol{a} \cdot \boldsymbol{v})MW}{N\sqrt{\lambda f}}.
\end{equation} The error of the received output field could be corrected by shaping the field after the slit, but this would require knowing the expected output field before performing each optical dot product apply the appropriate correction mask. This, of course, defeats the purpose of the optical computing system, since correcting the output field would require that we know the output field and thus also the dot product solution. Therefore, wave optics dictates that some error will inherently be present in $2f$ MVM, even in the absence of noise.

\subsection{4f MVM}
The $4f$ architecture implements the following method for performing the matrix-vector product $\boldsymbol{A}\boldsymbol{v}$, described in detail in Method 2 of the Supplementary Information:
\begin{enumerate}
    \item The cross-correlation is calculated between vector $\boldsymbol{v}$ and each row of $\boldsymbol{A}$. Equivalently, $\boldsymbol{v}$ can be reversed and then convolved with each matrix row.
    \item The "central" component of each resulting cross-correlation encodes the dot product between $\boldsymbol{v}$ and each row of $\boldsymbol{A}$. The vector of central components across all matrix rows in the solution $\boldsymbol{A}\boldsymbol{v}$. 
\end{enumerate}

The $4f$ architecture follows the $4f$ correlator setup described in \cite{Heinz} (see Fig.~\ref{fig:2f_4f_mvm}(B)). The input field $U_{\mathrm{in}}$ encodes the vector $\boldsymbol{v}$ similarly to the $2f$ architecture, except that spaces are inserted between each vector element to prevent peaks from overlapping in the final output field. The first cylindrical lens in the setup maps $U_{\mathrm{in}}$ in the $x_1y_1$-\nobreakdash plane to its Fourier transform (along the $x$-dimension) in $x_2y_2$-\nobreakdash plane. The modulator applies the transmittance function $t_{4f}$ in the $x_2y_2$-\nobreakdash plane, modulating the angular spectrum of $U_{\mathrm{in}}$. The applied transmittance function $t_{4f}$ is the sampled Fourier transform (along the $x$-dimension) of a function $A(-x, y)$, where $A(x, y)$ "directly" encodes the matrix $\boldsymbol{A}$. We say that the modulator transmittance "indirectly" encodes $\boldsymbol{a}$ because it is the Fourier transform of the function $A$ whose pulse amplitudes exactly ("directly") correspond to the elements $a_n$. The second cylindrical lens performs a Fourier transform of the modulated field, yielding the output field $U_{\mathrm{out}}(x, y) = U_{\mathrm{in}} \, (-x, y) (\ast_x) \, A(x, y)$, which is the convolution along the $x$-dimension of the "reversed" input field $U_{\mathrm{in}}(-x, y)$ and the function $A(x, y)$, which serves as a convolution kernel. If the functions $A$ and $U_{\mathrm{in}}$ are pulse trains with sufficient spacing between each pulse, then the output field is composed of an array of scaled peaks. A slit selects out the desired convolution peaks which encode $\boldsymbol{A}\boldsymbol{v}$ in the output field.  

\subsubsection{Analytic Calculations: 4f Architecture} \label{subsec:4f_MVM_analytic}
We use scalar wave optics theory to calculate the evolution of a horizontal $x$-\nobreakdash cross section of the input field $U_{\mathrm{in}}$ through a $4f$ MVM system computing $D = \boldsymbol{a} \cdot \boldsymbol{v}$. We refer to Fig.~\ref{fig:2f_4f_mvm}(B) for all coordinate axes labels and function names, but we allow the focal length of the first cylindrical lens, $f_1$, to differ from that of the second lens, $f_2$. 

We define a function $A(x)$ that encodes $\boldsymbol{a}$, composed of rectangular pulses scaled by vector elements $a_n$:
\begin{equation}\label{eq:A(x)}
    A(x) = \sum_{n=1}^N a_n \rect \! \left(\frac{x - ns - \psi_0}{T} \right),
\end{equation}
where $T$ is the width of the rectangular pulses, $s$ is the spacing between pulses, and $\psi_0$ is a centering parameter. The input optical field $U_{\mathrm{in}}$ encodes vector $\boldsymbol{v}$ in the same way:
\begin{equation}\label{eq:U_in,4f}
    U_{\mathrm{in}}(x) = \sum_{n=1}^N v_n \rect \! \left(\frac{x - ns - \psi_0}{T} \right).
\end{equation}
The $4f$ architecture effectively performs $U_{\mathrm{in}}(-x) \ast A(x)$, so we must have $s \geq 2T$ to prevent convolution peaks from overlapping in the output field.  

We pattern the Fourier transform of $A(-x)$ onto the modulator transmittance function $t_{4f}$ according to 
\begin{equation}\label{eq:modulator_t_4f}
    t_{4f}(x_2) = \sum_{m=1}^M t_m \rect \! \left(\frac{x_2 - mr - X_0}{W} \right),
\end{equation}
where 
\begin{equation}
    \begin{aligned}\label{eq:fourier_of_vec_a}
        t_m & = \int_{-\infty}^{\infty} \mathcal{F}_x\{ A(-x_1) \}(\nu) \: \delta(\nu - m\Omega_s - \omega_0) \, \mathrm{d}\nu\\
        & = \int_{-\infty}^{\infty} \int_{-\infty}^{\infty} A(-x_1) \, \exp\!{(-i2\pi \nu x_1)} \, \delta(\nu - m\Omega_s - \omega_0) \, \mathrm{d}x_1 \, \mathrm{d}\nu \\
        & = \int_{-\infty}^{\infty} A(-x_1) \, \exp\!{\left [-i2 \pi (m\Omega_s + \omega_0)x_1 \right ]} \, \mathrm{d}x_1
    \end{aligned}
\end{equation}
are the samples of the Fourier transform. The modulator is composed of $M$ rectangular pixels of full-width $W$, spaced at a pitch of $r$, and we introduce a centering offset $X_0$ to the modulator transmittance function. Although we wish to program $t_{4f}$ to encode the Fourier transform of $A(-x)$, since the modulator is composed of finitely-many rectangular pixels, we can do so only approximately. Therefore, the transmission coefficients $t_m$ are evenly-spaced samples of the spectrum of $A(-x)$ at frequencies $\nu_m = m\Omega_s + \omega_0$, where $\Omega_s$ is the sampling period in spatial frequency space and $\omega_0$ is an offset. 

Having determined $t_{4f}$ and $U_{\mathrm{in}}$, we borrow the result from Eq.~(S16) (see Supplementary Information) for the $4f$  output field
\begin{equation} \label{eq:4f_convolution}
    \begin{aligned}
        U_{\mathrm{out}, 4f}(x_3) &  = \sqrt{\frac{f_1}{f_2}} \, \big[ U_{\mathrm{in}}(-x_1) * \mathcal{F}_\nu \{ t_{4f}(\lambda f_1\nu) \}(x_1) \big] \! \left( \frac{f_1 x_3}{f_2} \right) \\
        & = \sqrt{\frac{f_1}{f_2}} \, \big[ U_{\mathrm{in}}(-x_1) * A_{\mathrm{eff}}(x_1) \big] \! \left(\frac{f_1 x_3}{f_2}\right). 
    \end{aligned}
\end{equation}
where $A_{\mathrm{eff}}$ is the "effective $A(x)$" encoded by $t_{4f}$ when accounting for the sampling period and the shape, width, spacing, and number of pixels. $A_{\mathrm{eff}}$ acts as a convolution kernel representing the impulse response of the $4f$ MVM system. We calculate that
\begin{equation}
    \begin{aligned}
        A_{\mathrm{eff}}(x_1) & = \mathcal{F}_x\{ t_{4f} \}\! \!\left(\frac{x_1}{\lambda f_1} \right) = \mathcal{F}_x\{ t_{4f} \}(\nu);\quad \nu = \frac{x_1}{\lambda f_1}\\
        & =
        \frac{W}{\lambda f_1} \sinc(\pi W \nu) \exp\!{(-i2\pi X_0 \nu)}  \sum_{m=1}^{M} t_m \exp\!{(-i2 \pi mr \nu)} \\ 
        & = \begin{multlined}[t]
        \frac{W}{\lambda f_1} \sinc(\pi W \nu) \exp\!{(-i2\pi X_0 \nu)} \\
        \times \int_{-\infty}^{\infty} A(-\xi) \, \exp\!{(-i2\pi \omega_0 \xi)} \,
        \sum_{m=1}^{M} \exp\!{\left[-i2 \pi m(\Omega_s \xi +  r\nu)\right]} \, \mathrm{d}\xi \end{multlined}
        \\
        & = \begin{multlined}[t]
        \frac{W}{\lambda f_1} \sinc(\pi W \nu)\exp\!{(-i2\pi X_0 \nu)} \\ \times \int_{-\infty}^{\infty} A(-\xi) \, \exp\!{(-i2\pi \omega_0 \xi)} \, \exp\!{\left [-i\pi (M+1)(\Omega_s \xi + r \nu)\right ]} \, \\ \times \frac{\sin\! \left [\pi M (\Omega_s \xi + r \nu)\right ]}{\sin\! \left [\pi(\Omega_s  \xi + r\nu) \right ]} \, \mathrm{d}\xi
        \end{multlined}
    \end{aligned}
\end{equation}
and choosing $X_0 = -(M+1)r/{2}$ and $\omega_0 = -(M+1)\Omega_s/{2}$, we obtain
\begin{equation} \label{eq:A_eff_analytic_prediction}
    \begin{aligned}
        A_{\mathrm{eff}}(x_1) & = \frac{W}{\lambda f_1} \sinc(\pi W \nu) \int_{-\infty}^{\infty} A(-\xi) \, \frac{\sin\! \left [\pi M (\Omega_s \xi + r \nu) \right ]}{\sin \!\left [\pi(\Omega_s  \xi + r\nu) \right ]} \, \mathrm{d}\xi \\
       & = \frac{W}{\lambda f_1} \sinc(\pi W \nu) \int_{-\infty}^{\infty} A(\beta) \, \frac{\sin \!\left [\pi M \Omega_s( r \nu / \Omega_s - \beta ) \right ]}{\sin \!\left [\pi \Omega_s( r \nu /\Omega_s - \beta) \right]} \, \mathrm{d}\beta.
    \end{aligned}
\end{equation}
From this we see 
\begin{equation} \label{eq:final_A_eff}
A_{\mathrm{eff}}(x_1) = \frac{W}{\lambda f_1} \sinc \! \left (\frac{\pi W x_1}{\lambda f_1} \right ) \left(A \ast D \right) \! \left(\frac{r x_1}{\lambda f_1 \Omega_s}\right )
\end{equation}
where $D(x_1) = \sin(\pi M \Omega_s x_1)/{\sin(\pi \Omega_s x_1)}$. Fixing the wavelength of the light source $\lambda$ and $M$, $W$, and $r$ for the modulator, we choose $\Omega_s = r/{(\lambda f_1)}$. Therefore, sampling at the proper period $\Omega_s$ to bring $A_{\mathrm{eff}}(x_1)$ as "close" to $A(x_1)$ as possible requires choosing an optimal focal length $f_1$ for the first cylindrical lens. Alternatively, a $4f$ magnifying system may be placed between the first cylindrical lens and the modulator to optimize frequency sampling. After substituting $\Omega_s = r/(\lambda f_1)$ and $W = F \, r$, where $F$ is the modulator pixel fill-factor, we obtain 
\begin{equation} \label{eq:final_A_eff}
A_{\mathrm{eff}}(x_1) = F\, \Omega_s\, \sinc(\pi F \Omega_s x_1) \, (A \ast D)(x_1).
\end{equation}
For fixed fill-factor $F$, we vary $\Omega_s$ and $M$ to control how closely $A_{\mathrm{eff}}$ approximates $A$ (up to a constant factor). The Dirichlet function $D(x)$ has peaks at $x = n/{\Omega_s}$ of value $M$ for even $n$ and $-M$ for odd $n$. To prevent aliasing when $A$ is convolved with $D$, we must have
\begin{equation}\label{eq:samp_frequency}
    \frac{1}{\Omega_s} \geq L_A = (N-1)s + T
\end{equation}
where $L_A$ is the spatial extent of $A$. We observe that $D$ approaches the desired form of a delta function under
\begin{equation}
    \lim_{M \Omega_s \to \infty}  D(x_1) = M \,  \delta(x_1).
\end{equation}
Any other shape than a perfect delta function will introduce "blurring" due to imperfect imaging. Also, to prevent distortion from the $\sinc$ envelope, $\Omega_s$ must be small. Therefore, we must have $\Omega_s$ small enough to prevent aliasing and distortion but $M \Omega_s$ large enough to prevent excessive blurring when $A$ is convolved with $D$. Therefore, there must be an inverse relationship between $M$ and $\Omega_s$ to keep the degree of blurring constant, so, by Eq.~(\ref{eq:samp_frequency}), $M$ must grow linearly with $N$ to accurately encode $A$ as vector length grows. 

We plug Eq.~(\ref{eq:final_A_eff}) into Eq.~(\ref{eq:4f_convolution}) to obtain the output field. Since $A(x)$ and $U_o(x)$ are composed of rectangular pulses, the output field is composed of an array of triangular pulses. A slit selects out the desired central triangular pulse encoding the dot product $D = \boldsymbol{a} \cdot \boldsymbol{v}$.

\subsubsection{4f Dot Product Example}\label{subsec:4f_example}

\begin{figure*}[htbp]
  \noindent
  \centering {\includegraphics[width=0.7\textwidth]{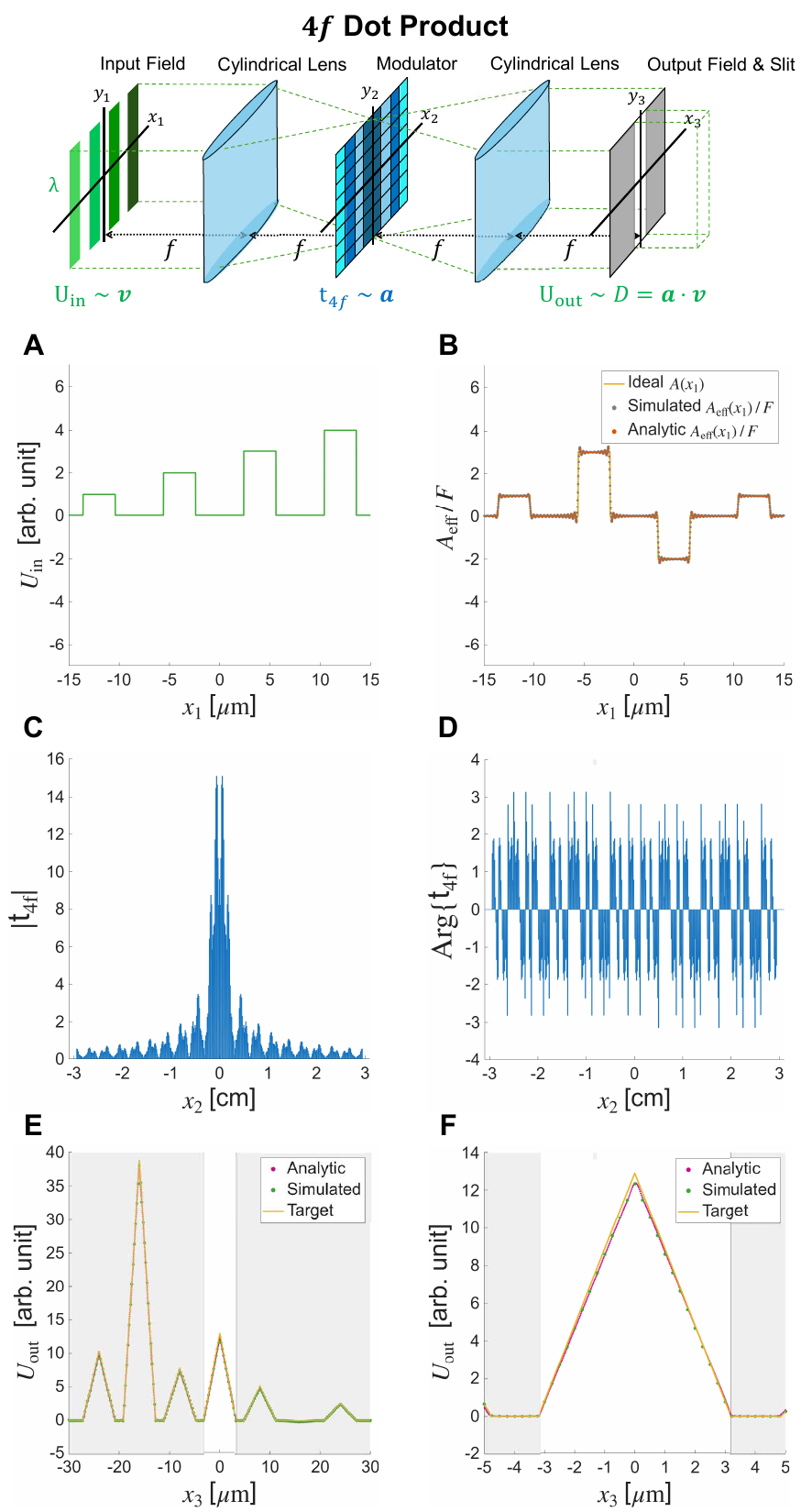}} \caption{\label{fig:4f_dot_prod_fields_example}
  Top: Diagram of $4f$ architecture performing vector dot product $D = \boldsymbol{a} \cdot \boldsymbol{v}$ showing input field $U_{\mathrm{in}}$, modulator transmittance $t_{4f}$, and output field $U_{\mathrm{out}}$ and relevant planes.
  (A) Input optical field $U_{\mathrm{in}}$ encoding vector $\boldsymbol{v} = (1, 2, 3, 4)^T$. (B) Normalized convolution kernel $A_{\mathrm{eff}}/F$, where $F$ is the modulator pixel fill factor, derived from the modulator transmittance of (C) and (D). The numerically simulated kernel (gray) confirms that analytically predicted (dark orange) by Eq.~(\ref{eq:A_eff_analytic_prediction}). Comparison is made to the ideal kernel $A(x)$ (light orange) "directly" encoding $\boldsymbol{a} = (1, 3, -2, 1)^T$. (C) Magnitude of the modulator transmittance function $t_{4f}$ residing in the $x_2y_2$-\nobreakdash plane, "indirectly" encoding vector $\boldsymbol{a}$. The comb-like appearance of the plot trace is due to the $M = 401$ individual pixels of the modulator. (D) Phase of the modulator transmittance function $t_{4f}$. (E) Output field $U_{\mathrm{out}, 4f}$ in the slit plane $x_3y_3$ encoding solution $D = \boldsymbol{a} \cdot \boldsymbol{v}$, compared to the desired target field (light orange). Simulated results (green) confirm the output field predicted analytically (magenta) from Eq.~(\ref{eq:4f_convolution}). The portion of the field not passed by the slit is grayed out. (F) Zoomed in version of (E). The slit width is $\Delta x_3 = 6.4 \, \mathrm{\mu m}$. }
  \end{figure*}

As we did for the $2f$ architecture, we consider the $4f$ architecture performing the dot product of two $N = 4$ length vectors $D = \boldsymbol{a}  \cdot  \boldsymbol{v} = (1, 3, -2, 1)^T \cdot (1, 2, 3, 4)^T = 5$, where $\boldsymbol{v}$ is encoded in the optical field $U_{\mathrm{in}}$ and $\boldsymbol{a}$ is "indirectly" encoded in the modulator transmittance function $t_{4f}$ and "directly" encoded in the convolution kernel $A_{\mathrm{eff}}$ (see Fig.~\ref{fig:4f_dot_prod_fields_example}(A--D)). In this example, the pulses of the input field and convolution kernel have width $T = 3.2 \, \mathrm{\mu m}$ and their centers are spaced at $s = 8 \, \mathrm{\mu m}$, so the total spatial extent of the kernel is $L_A = 27.2 \, \mathrm{\mu m}$ by Eq.~(\ref{eq:samp_frequency}). We choose focal lengths $f = f_1 = f_2 = 2 \, \mathrm{cm}$ and use light of wavelength $\lambda = 0.5 \, \mathrm{\mu m}$. We sample the Fourier transform of $A$ in spatial frequency space at period $\Omega_s = 0.4/L_A \approx 0.0147 \, \mathrm{\mu m}^{-1}$, so the real-space modulator pixel pitch is $r = \lambda f \Omega_s \approx 0.0147 \, \mathrm{cm}$. With modulator pixel fill-factor of $F = 17/21$, the pixel width is $W = Fr \approx 0.0119 \, \mathrm{cm}$. The modulator has a total of $M=401$ pixels, giving a total length $L_{\mathrm{mod}} \approx 5.892 \, \mathrm{cm}$ wide. 

After patterning the modulator transmittance function, we find how well the modulator encodes vector $\boldsymbol{a}$ by examining how closely $A_{\mathrm{eff}}(x)$ resembles the ideal kernel $A(x)$. As shown in Fig.~\ref{fig:4f_dot_prod_fields_example}(B), the finite number of modulator pixels introduces wiggles which cause $A_{\mathrm{eff}}$ to deviate from $A$. The non-unity fill-factor $F$ causes optical energy to be lost to the modulator, so the plotted $A_{\mathrm{eff}}$ curves are divided by $F$ to allow comparison of the shapes of $A$ and $A_{\mathrm{eff}}$. The numerically simulated $A_{\mathrm{eff}}$, found by applying FFT to $t_{4f}$, confirms the analytic result of Eq.~(\ref{eq:final_A_eff}). 

The corresponding output field $U_{\mathrm{out}, 4f}$ is shown in Fig.~\ref{fig:4f_dot_prod_fields_example}(E) and (F). The output slit selects the central peak carrying the dot product solution within the interval $x_3 \in [-T , T]$. The numerically simulated output field confirms the output field predicted analytically from Eq.~(\ref{eq:4f_convolution}) and Eq.~(\ref{eq:final_A_eff}). The target field within the slit is a perfectly triangular waveform 
\begin{equation}\label{eq:target_U_out_4f}
    U_{\mathrm{out}, 4f}^{\mathrm{target}}(x_3) = \frac{(\boldsymbol{a} \cdot \boldsymbol{v})TF}{\mu \mathrm{m}} \,  \Lambda(x_3/T) = 12.8 \,  \Lambda(x_3/T),
\end{equation} where \begin{equation*}
    \Lambda(x) = \begin{cases}
    1 - |x|, & |x| < 1 \\
    0, & |x| \geq 1 
    \end{cases}
\end{equation*}
is a triangle function. The desired dot product is encoded in the amplitude of this central triangular peak. As shown in Fig.~\ref{fig:4f_dot_prod_fields_example}(F), the actual received output field deviates from this perfectly triangular form. If the modulator had infinitely many pixels that were infinitesimal in width, the central peak would have a perfect triangular shape for any dot product (of real-valued vectors) performed. This is in contrast to the $2f$ architecture, where the shape of the field within the slit depends on the dot product performed. Without infinitely many modulator pixels, some error will always be present in the computed $4f$ dot product solution since $A_{\mathrm{eff}}$ will deviate from $A$ and therefore the output field will not be perfectly triangular. 

\section{Definitions}\label{sec:definitions}
We now define two quantities, called "waveform error" and "effective attenuation", to quantify and compare the performance of the $2f$ and $4f$ architectures. We plot these quantities in the Results Sec.~\ref{sec:results}.

We have seen that, for both $2f$ and $4f$ architectures, the actual received output waveform $U^{\mathrm{actual}}_{\mathrm{out}}$ will deviate from the target waveform $U^{\mathrm{target}}_{\mathrm{out}}$. This introduces error in the optically-computed dot product or MVM. We therefore define "percent waveform error" $\mathrm{WE}$, which measures the degree to which an output waveform encoding the dot product solution differs from the corresponding target waveform within the output slit:
\begin{equation}\label{eq:waveform_err_def}
    \mathrm{WE} = 100  \left ( \frac{\int_{\mathrm{slit}} \left|U^{\mathrm{target}}_{\mathrm{out}} - U^{\mathrm{actual}}_{\mathrm{out}}\right| \, \mathrm{d}x}{\int_{\mathrm{slit}} \left|U^{\mathrm{target}}_{\mathrm{out}}\right| \, \mathrm{d}x} \right ).
\end{equation}

Since the slits in the $2f$ and $4f$ output planes select only small slivers of the output fields, we should expect that only a fraction of the optical energy of the input field should exit an MVM system encoding the dot product solution. If we wish to perform cascaded MVM, where the output of a matrix-vector product is multiplied by another matrix, then we may need to amplify the received signal before it enters another MVM system. We therefore wish to quantify the attenuation of an optical signal after passing through an MVM system. Suppose that an input field $U_{\mathrm{in}}$ encoding $\boldsymbol{v}$ enters an optical dot product system. One pulse of $U_{\mathrm{in}}$ encodes the numerical value $v_1 = 1$ and the pulse has optical energy $E_1$. After the field passes through the system, suppose that the part of the output field exiting the slit encodes numerical value $D  = \boldsymbol{a} \cdot \boldsymbol{v}= 1$ and has optical energy $E_1'$. We define a quantity $\alpha_{\mathrm{eff}}$, called "effective attenuation", that quantifies the ratio of the energy of an output signal pulse to the energy of an input signal pulse, when the output and input pulses encode the same numerical value. After amplifying the output signal by the inverse of $\alpha_{\mathrm{eff}}$, an output pulse encoding numerical value $c$ will have the same energy as an input pulse encoding $c$. Effective attenuation $\alpha_{\mathrm{eff}}$ measures signal attenuation, when accounting for the numerical values encoded by the field entering and exiting an MVM system. In the above example of input and output pulses encoding numerical value $1$, the effective attenuation would be $\alpha_{\mathrm{eff}} = E_1'/E_1$. A smaller effective attenuation factor means that an optical signal experiences more loss after being processed in an optical MVM system. For an optically-computed dot product $D = \boldsymbol{a} \cdot \boldsymbol{v}$, we define "effective attenuation" as 
\begin{equation}\label{eq:define:eff_atten}
    \alpha_{\mathrm{eff}} = \frac{\left|\boldsymbol{v}\right|^2 E_{\mathrm{out}}}{D^2 \, E_{\mathrm{in}}} = \frac{\left|\boldsymbol{v}\right|^2 \int_{\mathrm{slit}}\left|U_{\mathrm{out}}\right|^2 \, \mathrm{d}x}{D^2  \int_{-\infty}^{\infty}\left|U_{\mathrm{in}}\right|^2 \, \mathrm{d}x},
\end{equation}
where $E_{\mathrm{in}}$ is the total optical energy of the input field $U_{\mathrm{in}}$ and $E_{\mathrm{out}}$ is the optical energy of the output field $U_{\mathrm{out}}$ within the slit. In the formula, we normalize these energies so that the factor $E_{\mathrm{in}}/|\boldsymbol{v}|^2$ is the energy of an input pulse encoding numerical value $1$ and $E_{\mathrm{out}}/D^2$ is the energy of an output pulse encoding $1$. In general, $\alpha_{\mathrm{eff}}$ may vary for different dot products performed. To estimate "effective attenuation per MVM", plotted in the Results Sec.~\ref{sec:results}, we may take the average $\alpha_{\mathrm{eff}}$ over many dot products. Since MVM consists of many dot products performed in parallel, averaging over many dot products properly extends the definition of effective attenuation to full MVM. 

\section{Methods}

Below, we discuss the computational models used to simulate the $2f$ and $4f$ architectures. First, we discuss the model used to simulate a realistic modulator with a maximum optical gain limit and finite bit depth. Understanding our modulator model is essential for understanding the scaling of effective attenuation per MVM (defined in Sec.~\ref{sec:definitions}) with MVM problem size for the $2f$ and $4f$ architectures. Then, we discuss the assumptions and computational steps of our simulation and list the parameter sweeps we performed to obtain our results.

\subsection{Modeling the Modulator}\label{subsec:modeling_the_modulator}

\begin{figure}[htbp]
  \noindent 
  \centering {\includegraphics[width=0.9\textwidth]{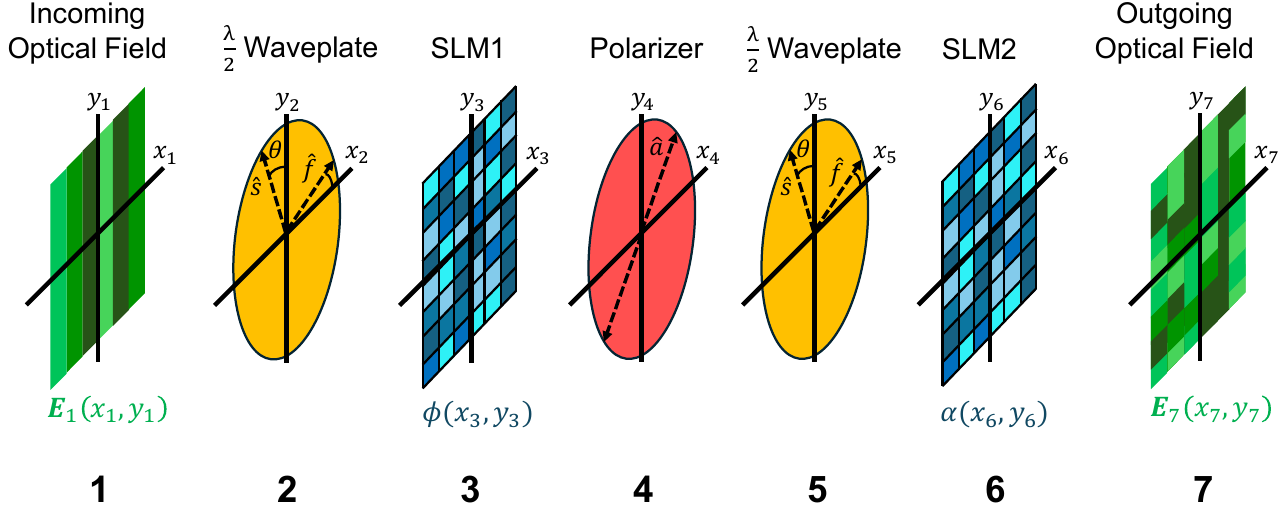}} \caption{\label{fig:SLMs_polarizers} Schematic of a free-space modulator capable of independently manipulating the intensity and phase of incoming light. Numbers denote steps to manipulate the magnitude and phase of an incoming electric field. (1) $\hat{x}$\nobreakdash-polarized electric field $\boldsymbol{E}_1 = U(x_1, y_1) \hat{x}$ enters from the left. (2) $\lambda/2$ waveplate has fast and slow axes $\hat{f}$ and $\hat{s}$ which are rotated $\theta = \frac{\pi}{8}$ counterclockwise from the $x$ and $y$-axes, respectively.(3) SLM1 applies a phase $\phi(x_3, y_3)$ to the $\hat{x}$\nobreakdash-polarized component of the field relative to the $\hat{y}$\nobreakdash-polarized component. (4) Polarizer with axis of polarization $\hat{a} = \frac{\hat{x} + \hat{y}}{\sqrt{2}}$ reduces the intensity of the field dependent on the $\phi$ phase profile. (5) $\lambda/2$ waveplate has fast and slow axes $\hat{f}$ and $\hat{s}$ which are rotated $\theta = \frac{\pi}{8}$ counterclockwise from the $x$ and $y$-axes, respectively. (6) SLM2 applies the phase profile $\alpha(x_6, y_6)$ to the $\hat{x}$\nobreakdash-polarized component of the field relative to the $\hat{y}$\nobreakdash-polarized component. (7) $\hat{x}$\nobreakdash-polarized electric field $\boldsymbol{E}_7$ exits to the right. $\phi$ controls its intensity while $\alpha$ controls its phase. }
\end{figure}

Until this point, we have imagined the modulator as a single infinitesimally thin device whose pixels can be tuned to encode any complex-valued transmission coefficient. In the lab, two phase-only spatial light modulators (SLMs) can be used to achieve reprogrammable intensity and phase modulation in free space. We construct a desired transmittance function $t(x, y)$ from two phase-only SLMs by following the methods of \cite{zhu2014arbitrary}. Referring to the optical system in Fig.~\ref{fig:SLMs_polarizers}, we walk through the steps below:
\begin{enumerate}
    \item An electric field $\boldsymbol{E}_1(x_1, y_1) = U(x_1,y_1)\hat{x}$, corresponding to optical field $U$, enters the system polarized in the $\hat{x}$\nobreakdash -direction. 
    \item A $\lambda/2$ waveplate, whose fast $\hat{f}$ and slow $\hat{s}$ axes are rotated $\theta = \frac{\pi}{8}$ radians counterclockwise from the $x$ and $y$-axes, respectively, rotates the polarization of the field according to $\hat{x} \to \left[(\hat{x} + \hat{y})/\sqrt{2}\right]$. This yields the field immediately after the waveplate $\boldsymbol{E}_2(x_2, y_2) = U(x_2,y_2)\left[(\hat{x} + \hat{y})/\sqrt{2}\right]$. 
    \item The phase-only SLM1 applies the phase profile $\phi(x_3, y_3)$ to the $\hat{x}$\nobreakdash-polarized component of the field relative to the $\hat{y}$\nobreakdash-polarized component, yielding $\\ \boldsymbol{E}_3(x_3, y_3) = U(x_3,y_3)\{\exp\!{[i\phi(x_3, y_3)]}\hat{x} + \hat{y}\}/\sqrt{2}$. 
    \item The field then passes through a polarizer with axis of polarization $\hat{a} = (\hat{x} + \hat{y})/{\sqrt{2}}$, yielding $\boldsymbol{E}_4(x_4, y_4) = U(x_4,y_4)\cos[{\phi(x_4, y_4})]\exp\!{[i\phi(x_4, y_4)/2]}\,\hat{a}$.
    \item A second $\lambda/2$ waveplate, whose fast and slow axes are oriented exactly as the other waveplate, rotates the polarization according to $\hat{a} \to \hat{x}$, yielding $\\ \boldsymbol{E}_5(x_5, y_5) = U(x_5,y_5)\cos\!{[\phi(x_5, y_5)]}\exp\!{[i\phi(x_5, y_5)/2]}\, \hat{x}$. 
    \item The phase-only SLM2 applies the phase profile $\alpha(x_6, y_6)$ to the $\hat{x}$\nobreakdash-polarized component of the field relative to the $\hat{y}$\nobreakdash-polarized component, yielding $\\ \boldsymbol{E}_6(x_6, y_6) = U(x_6,y_6)\cos\!{[\phi(x_6, y_6)]}\exp\!{\{i[\phi(x_6, y_6) + 2\alpha(x_6, y_6)]/2\}} \, \hat{x}$.
    \item The field exiting the system is $\\ \boldsymbol{E}_7(x_7, y_7) = U(x_7,y_7)\cos\!{[\phi(x_7, y_7)]}\exp\!{\{i[\phi(x_7, y_7) + 2\alpha(x_7, y_7)]/2\}} \, \hat{x}$, where we see that the phases $\phi$ and $\alpha$ applied at the SLMs provide independent control over the magnitude and phase profiles, respectively, of the optical field at the output. 
\end{enumerate}
Passage through the above system is equivalent to passing through an infinitely-thin "effective modulator" with transmittance function
\begin{equation}\label{eq:modulation_states}
    t(x, y) = \frac{\boldsymbol{E}_7(x, y)}{\boldsymbol{E}_1(x, y)} = \cos \! \left[{\phi(x, y})\right]  \exp\!\left\{\frac{i}{2}\big[\phi(x,y) + 2\alpha(x,y)\big]\right\}.
\end{equation}
If we allow the phases $\phi$ and $\alpha$ to occupy any value on the interval $[-\pi, \pi]$, then we can program the effective modulator pixel of the $m$th row and $n$th column, $t_{mn}$, to encode any transmission coefficient within the unit complex disk. However, we model our SLMs to have finite bit depth $b \in \mathbb{Z}^+$, so that each SLM can only access discrete phase levels from the set 
\begin{equation}
    \Phi = \left\{ \frac{2\pi k}{2^b} : k = 0, 1, \ldots, 2^b - 1 \right\}.
\end{equation}
Knowing the phases of $\phi$ and $\alpha$ must be elements of set $\Phi$, we use Eq.~(\ref{eq:modulation_states}) to determine the set, $\tau$, of possible values that the transmittance function can take:
\begin{equation}\label{eq:tau_phase_states}
    \tau =  \left\{ \cos(\phi)  \exp \! {\left[ \frac{i}{2}(\phi + 2\alpha)\right]} : \phi, \alpha \in \Phi \right\}.
\end{equation}

In our simulations, we assumed that both SLMs had the same pixel dimensions, pitch, and fill-factor and that perfect imaging occurred between the two SLM planes. We were therefore able to  model the modulator system as an infinitesimally thin effective modulator which shared the same pixel dimensions, pitch, and fill-factor and whose transmittance function took values from $\tau$. To approximately encode an ideal transmittance function $t^{\mathrm{ideal}}$ on the modulator, we first sampled the ideal transmittance value at each $m$th modulator pixel, $t^{\mathrm{ideal}}_m$. Since Eq.~(\ref{eq:modulation_states}) dictates $|t(x, y)| \leq 1$ and we chose to use the full dynamic range of the modulator, we then divided each of the ideal transmittance samples by the magnitude of the largest sample, yielding $t^{\mathrm{norm}}_m = t^{\mathrm{ideal}}_m/\left(\max_{n}\left|t^{\mathrm{ideal}}_n\right|\right)$ for the $m$th pixel. This rescaling of the ideal modulator transmittance function greatly impacts how effective attenuation per MVM scales with MVM problem size, as predicted in Sec.~\ref{subsec:scalings_from_stats} and seen in the Results Sec.~\ref{sec:results}. Note that the part of the output field within the slit may be intentionally boosted by $\max_{n} |t^{\mathrm{ideal}}_n|$ to correct for this rescaling factor after performing an optical dot product. Finally, we chose the value $t \in \tau$ closest to $t^{\mathrm{norm}}_m $ that minimizes $\left|t - t^{\mathrm{norm}}_m \right|$. The transmittance of the $m$th pixel was assigned $t_m = t$. We repeated this for each modulator pixel to determine the effective transmittance function corresponding to two phase-only SLMs of finite bit depth $b$. 

\subsection{Constructing Our Simulations}
We simulated propagation of an input optical field $U_{\mathrm{in}}$ through the $2f$ and $4f$ MVM systems using Fourier optics. A complex field $U(x)$ represents a (unitless) scalar electric field amplitude with $|U(x)|^2$ proportional to intensity. We modeled a lens to perform an exact Fourier transform between its front and back focal planes, assuming the pupil function to be unity across all space. The evolution of an optical field through the $2f$ and $4f$ systems was simulated by sequentially computing the field at each (front and back) focal plane. Our simulations tracked the evolution of a horizontal $x$-cross section of the input field through the systems. Therefore, we simulated $2f$ and $4f$ dot products, as explored analytically in Sec.~\ref{subsec:2f_MVM_analytic} and Sec.~\ref{subsec:4f_MVM_analytic}. The input field and modulator transmittance function were sampled in one dimension, and a 1D Fast Fourier Transform (FFT) was sequentially applied to determine the sampled fields in each focal plane. We modeled the modulator as composed of two 8\nobreakdash-bit ($b = 8$) SLMs and determined the set of programmable transmittance values from Eq.~(\ref{eq:tau_phase_states}). We further assumed no reflection losses at the lenses or modulator interfaces. In the following, the $2f$ and $4f$ field and transmittance functions are constructed as in Sec.~\ref{sec:2f_and_4f_MVM}.

For the $2f$ simulation, we created a 1D rectangular pulse motif representing a pixel of the modulator transmittance function. The pulse motif was 21 samples wide, composed of two samples at value 0, seventeen samples at value 1, and two samples at value 0, giving a modulator pixel fill-factor of $F = 17/21$. The samples were spaced at $\Delta x_1 = (1.6 \, \mathrm{cm})/(21 M)$, where $L_{\mathrm{mod}} = 1.6 \, \mathrm{cm}$ is the full-width of the modulator and $M$ is the modulator pixel count. For the $m$th modulator pixel, we determined the ideal transmittance value $t_m^{\mathrm{ideal}}$ and substituted the corresponding transmittance level $t_m \in \tau$ (refer to Sec.~\ref{subsec:modeling_the_modulator} for more details). The rectangular pulse motifs were repeatedly scaled by the corresponding values $t_m$ and concatenated $M$ times to create the full modulator transmittance array. The input scalar optical field $U_{\mathrm{in}}$ was sampled at the same spacing, except its rectangular pulse motif was composed of 21 samples all at value 1. This pulse motif was repeated $M$ times and scaled appropriately to encode the amplitude profile of the $U_{\mathrm{in}}$ field.  This way, the field array was properly aligned with the modulator transmittance array. Zero-padding was applied identically to the modulator and field arrays to prevent artifacts when performing FFT. After calculating the output field array, we multiplied it by a windowing array to simulate a slit selecting spatial frequencies $\nu_x \in [-\Delta \nu_x/2, \Delta \nu_x/2]$ where $\nu_x = x_2/(\lambda f)$. 

For the $4f$ simulation, we chose $\lambda = 0.5 \, \mathrm{\mu m}$ and $f = f_1 = f_2 = 2 \, \mathrm{cm}$ and defined $A(x)$ and $U_{\mathrm{in}}(x)$ with spacing and rectangle width parameters $s = 4 \, \mathrm{\mu m}$ and $T = 3.2 \, \mathrm{\mu m}$ (see Eq.~(\ref{eq:A(x)}) and Eq.~(\ref{eq:U_in,4f})). We constructed a rectangular pulse motif representing a pixel of the modulator transmittance function that was 21 samples wide, composed of two samples at value 0, seventeen samples at value 1, and two samples at value 0 for a modulator pixel fill-factor of $F = 17/21$. This was identical to the $2f$ modulator. The pixel pitch was chosen as $r = \lambda f \Omega_s = \lambda f \left[ (0.4/L_A)(\mathrm{\mu m}^{-1}) \right ]$, where $N$ is the vector length and $L_A = [8(N -1) + 3.2] \, \mathrm{\mu m}$ is the spatial extent of $A$. Consequently, the spacing between samples in the modulator transmittance array was $\Delta x_2 = r/21$. We determined the ideal transmittance values to be $t^{\mathrm{ideal}}_m = \mathcal{F}_x\{ A \}(m r/\lambda f)$ for $-(M - 1)/2 \leq m \leq (M - 1)/2$, where $M$ is the modulator pixel count, and found the corresponding $t_m \in \tau$ according to the procedure in  Sec.~\ref{subsec:modeling_the_modulator}. We then scaled each $m$th pixel pulse motif by $t_m$ and concatenated them. Whitespace was added to the beginning and end of the scaled pulses to create the full modulator transmittance array of length $P$ samples (including whitespace). The input field array was created by sampling $U_{\mathrm{in}}(x_1)$ at $x_{1,p} = p/(\Delta x_2 \, P)$ for $-(P-1)/2 \leq p \leq (P-1)/2$. After calculating the output field array, we multiplied it by a windowing array to select the field over positions $x_3 \in [-T ,  T]$ within the slit. 

\subsection{Procedure}\label{sec:procedure}
Our simulations computed pairs of dot products $D=\boldsymbol{a} \cdot \boldsymbol{v}$ using the $2f$ and $4f$ architectures. We generated sets of vectors $\boldsymbol{a}$ and $\boldsymbol{v}$ at varying vector lengths $N \in \{25, 50, 100, 1000 \}$, where for each $N$, vector elements were sampled from three different statistical distributions. In the first "Gaussian" case of $N$\nobreakdash-length vectors, we sampled $a_n$ and $v_n$ from a Gaussian distribution so that $a_n, v_n \overset{\mathrm{i.i.d.}}{\sim} \mathcal{N}(0, 1)$ for $n=1, 2, ..., N$. In the second "sparse-Gaussian" case, we sampled $v_n$ from a Gaussian distribution so that $v_n \overset{\mathrm{i.i.d.}}{\sim} \mathcal{N}(0, 1)$ and $a_n$ from a sparse-Gaussian with $90\%$ sparsity so that
\begin{equation*}
    a_n = \begin{cases}
    0 & \text{with probability 0.9} \\
    \mathcal{N}(0, 1) & \text{with probability 0.1}
    \end{cases}
\end{equation*}
In the third "uniform" case, we sampled $a_n$ and $v_n$ from a uniform distribution so that $a_n, v_n \overset{\mathrm{i.i.d.}}{\sim} \mathcal{U}(0, 1)$. For each $N$ and each distribution ("Gaussian", "sparse-Gaussian", and "uniform"), we generated 1000 pairs of $N$\nobreakdash-length vectors $\boldsymbol{a}^{(t)},\boldsymbol{v}^{(t)}$ (where $t = 1, 2, ..., 1000$). 

For each pair of vectors $\boldsymbol{a}^{(t)}$ and $\boldsymbol{v}^{(t)}$, we simulated the $2f$ and $4f$ architectures performing the dot product $D^{(t)} = \boldsymbol{a}^{(t)} \cdot \boldsymbol{v}^{(t)}$. Our simulations numerically calculated the $2f$ and $4f$ output fields corresponding to each dot product, from which we determined the waveform error $\mathrm{WE}$ and effective attenuation $\alpha_{\mathrm{eff}}$ (defined in Sec.~\ref{sec:definitions}).

\subsection{Parameter Sweeps}
We first varied vector length $N$. For the $2f$ setup, we fixed slit bandwidth $\Delta \nu_x = 0.2 \, \mathrm{cm}^{-1}$ and varied $N$ over $N \in \{25, 50, 100, 1000\}$ at corresponding modulator pixel counts $M \in \{2000, 2000, 2000, 10000\}$. For the $4f$ setup, we varied $N$ over $N \in \{25, 50, 100, 1000\}$ at corresponding modulator pixel counts $M \in \{2001, 2001, 2001, 10001\}$. For each $N$, we numerically computed effective attenuations $\alpha_{2f}$ and $\alpha_{4f}$.

We then varied modulator pixel count $M$. For the $2f$ setup, we fixed slit bandwidth $\Delta \nu_x = 0.2 \, \mathrm{cm}^{-1}$ and vector length $N = 50$ and varied $M$ over $M \in \{400, 1000, 2000\}$. For the $4f$ setup, we fixed $N = 50$ and varied $M$ over $M \in \{401, 1001, 2001\}$. For each $M$, we numerically computed waveform errors $\mathrm{WE}_{2f}$ and $\mathrm{WE}_{4f}$ and effective attenuations $\alpha_{2f}$ and $\alpha_{4f}$.

We finally varied slit bandwidth $\Delta \nu_x$ for the $2f$ architecture. We fixed vector length $N = 50$ and modulator pixel count $M = 2000$ and varied slit bandwidth over $\Delta \nu_x \in \{0.05, 0.1, 0.2\}$. For each $\Delta \nu_x$, we numerically computed waveform errors $\mathrm{WE}_{2f}$ and effective  attenuation $\alpha_{2f}$.

\section{Predicted Attenuation Scalings}\label{sec:pred_scalings}
After specifying how we rescale and program the modulator transmittance function in Sec.~\ref{subsec:modeling_the_modulator}, we can now predict the scaling behavior of effective attenuation per MVM $\alpha_{\mathrm{eff}}$ with input vector length $N$. We approximate the effective attenuation by calculating the target output fields $U_{\mathrm{out}, 2f}$ and $U_{\mathrm{out}, 4f}$ and plugging into Eq.~(\ref{eq:define:eff_atten}). Although we calculate the scaling of $\alpha_{\mathrm{eff}}$ for arbitrary dot product $\boldsymbol{a} \cdot \boldsymbol{v}$ between $N$\nobreakdash-length vectors, the results are valid for full MVM as explained in the final paragraph of Sec.~\ref{sec:definitions} and later in this section.

We first examine the $2f$ case. Given the ideal $2f$ transmittance function
\begin{equation}\label{eq:modulator_t_2f}
    t^{\mathrm{ideal}}_{2f}(x_1) = \sum_{m=1}^{M} a_{k(m)} \rect \! \left (\frac{x_1 - mr - X_0}{W} \right);\quad  k(m) = \ceil \! \left(\frac{mN}{M} \right),
\end{equation}
and assuming we use the full dynamic range of the modulator and our modulator achieves a maximum amplitude gain of 1 (see Sec.~\ref{subsec:modeling_the_modulator} for more details), the desired target output waveform within the slit is 
\begin{equation}\label{eq:target_U_out,2f}
U^{\mathrm{target}}_{\mathrm{out}, 2f}(x) = \frac{(\boldsymbol{a} \cdot \boldsymbol{v})MW}{N\sqrt{\lambda f} \max\big|t^{\mathrm{ideal}}_{2f}(x)\big|} = \frac{DMW}{N\sqrt{\lambda f} \max_n|a_n|} \, .
\end{equation}
Above, we have used Eq.~(\ref{eq:target_U_out_2f}) and divided by the modulator transmittance rescaling factor $\max\big|t^{\mathrm{ideal}}_{2f}(x)\big|$ as explained in Sec.~\ref{subsec:modeling_the_modulator}. $D = \boldsymbol{a} \cdot \boldsymbol{v}$ is the dot product solution. Then, using Eq.~(\ref{eq:U_{in,2f}}) for $U_{\mathrm{in}}$ and the definition in Eq.~(\ref{eq:define:eff_atten}), we have 
\begin{equation}\label{eq:optical_energy_atten_2f}
     \alpha_{2f} =    \frac{\left|\boldsymbol{v}\right|^2 E_{\mathrm{out}, 2f}}{D^2 \, E_{\mathrm{in}, 2f}} \approx \frac{\Delta \nu_x F L_{\mathrm{mod}}}{N} \left( \max_n|a_n| \right)^{-2}, 
\end{equation}
where $\Delta \nu_x$ is the slit bandwidth, $F$ is the modulator pixel fill-factor, $L_{\mathrm{mod}}$ is the full-width of the modulator, and $N$ is the input vector length. This expression is approximate because we use the target output waveform rather than the actual received waveform.

In the $4f$ case, we sample the Fourier transform of $A(-x)$ at $x_2 = mr = m\lambda f \Omega_s$, for $-(M-1)/2 \leq m \leq (M-1)/2$ to find the value of the ideal transmittance function at the $m$th pixel:
\begin{equation}
    t^{\mathrm{ideal}}_{4f}(x_2 = m\lambda f \Omega_s) = \frac{T}{\sqrt{\lambda f}} \sinc(\pi T m\Omega_s) \sum_{n=1}^N a_n \exp \! {\big[ i2\pi (ns + \psi_o)m \Omega_s \big]}.
\end{equation}
Again assuming that we use the full dynamic range of the modulator and our modulator achieves a maximum amplitude gain of 1, we obtain the target output waveform
\begin{equation}\label{eq:target_U_out,4f}
U^{\mathrm{target}}_{\mathrm{out}, 4f}(x) = \left(  \frac{DF}{\max_{1 \le m \le M}
\left|
\sinc\!\big(\pi T m \Omega_s\big)
\sum_{n=1}^{N} a_n
\exp\!\big[i2\pi (ns+\psi_0)m\Omega_s\big]
\right|}  \right) \Lambda \! \left (\frac{x}{T} \right)
\end{equation}
within the slit. Above we have used Eq.~(\ref{eq:target_U_out_4f}) and divided by the modulator transmittance rescaling factor $\max\big|t^{\mathrm{ideal}}_{4f}(x)\big|$. Using Eq.~(\ref{eq:U_in,4f}) and Eq.~(\ref{eq:define:eff_atten}), we have 
\begin{equation}\label{eq:optical_energy_atten_4f}
     \alpha_{4f} =  \frac{\left|\boldsymbol{v}\right|^2 E_{\mathrm{out}, 4f}}{D^2 \, E_{\mathrm{in}, 4f}} \approx \frac{2 F^2}{3}\,
\left[
\max_{1 \le m \le M}\left|\sinc(\pi T m\Omega_s)\sum_{n=1}^{N}a_n\exp(i2\pi n m s\Omega_s)\right|\right]^{-2}.
\end{equation}

Although we have assumed that the maximum amplitude gain of the modulator is $1$, Eq.~(\ref{eq:optical_energy_atten_2f}) and Eq.~(\ref{eq:optical_energy_atten_4f}) can be modified to $G \alpha_{2f}$ and $G  \alpha_{4f}$ for a modulator of maximum amplitude gain $G$, assuming one uses the full dynamic range of the modulator. Note that these expressions depend on the particular $\boldsymbol{a}$ used in the dot product $D = \boldsymbol{a} \cdot \boldsymbol{v}$ and not on $\boldsymbol{v}$. We can predict the expected effective attenuation prior to optically computing a dot product and apply an appropriate signal boost to correct the attenuation. Also note that, even though our simulations compute effective attenuation for single dot products, effective attenuation can be extended to MVM, which consists of many dot products performed in parallel. For MVM, dot products between $\boldsymbol{v}$ and each row of a matrix $\boldsymbol{A}$ will generally have distinct effective attenuations. However, by averaging effective attenuation over many dot products, we can estimate the average "effective attenuation per MVM", which we plot in the Results Sec.~\ref{sec:results}.

\subsection{Scalings from Statistical Analysis}\label{subsec:scalings_from_stats}
We now predict the scaling of effective attenuation $\alpha_{2f}$ and $\alpha_{4f}$ with vector length $N$ when performing the dot product $D = \boldsymbol{a} \cdot  \boldsymbol{v}$. We predict attenuation scaling behaviors for different statistical distributions of the elements of $\boldsymbol{a}$. 

\subsubsection{2f Effective Attenuation Scalings}
From Eq.~(\ref{eq:optical_energy_atten_2f}) we have $\alpha_{2f} \propto N^{-1} \left[ \max_n|a_n| \right]^{-2}$. Suppose that all $a_n$, for $1 \leq n \leq N$, are drawn from a continuous random distribution with probability density function $f(a)$. Then the probability that a value $a$ is drawn whose magnitude is greater than some value $y$ is 
\begin{equation}
    \mathrm{P}(|a| > y) = \int_{y}^\infty f(|a|) \, \mathrm{d}a.
\end{equation}
For some value $a_{\mathrm{max}}$, we expect about one of $N$ total samples to be drawn with magnitude greater than $a_{\mathrm{max}}$ so that
\begin{equation} \label{eq:determine_a_max}
    N\,\mathrm{P}\!\left(|a| > a_{\mathrm{max}}\right) = 1.
\end{equation}
We solve this equality for $a_{\mathrm{max}}$ and estimate $\max_n|a_n| \approx a_{\mathrm{max}}$. 

Suppose the elements of $\boldsymbol{a}$ are Gaussian distributed: $a_n \overset{\mathrm{i.i.d.}}{\sim} \mathcal{N}(0, \sigma^2)$. Then $\\ f(|a|) = \sqrt{2/(\pi \sigma^2)} \exp[-a^2/(2 \sigma^2)]$ and 
\begin{equation}
\begin{aligned}
    \mathrm{P}(|a| > a_{\mathrm{max}}) &= \sqrt{\frac{2}{\pi \sigma^2}} \int_{a_{\mathrm{max}}}^{\infty} \exp\!{\left(\frac{-a^2}{2 \sigma^2} \right)} \, \mathrm{d}a \\
     & = \sqrt{\frac{2}{\pi \sigma^2}} \int_{0}^{\infty} \exp\!{\left(\frac{-(a + a_{\mathrm{max}})^2}{2 \sigma^2} \right)} \, \mathrm{d}a \\
     &\leq \exp\!{\left(\frac{-a_{\mathrm{max}}^2}{2 \sigma^2} \right)}. 
\end{aligned}
\end{equation}
Plugging this result into Eq.~(\ref{eq:determine_a_max}), we determine
\begin{equation}
    a_{\mathrm{max}} \sim \sqrt{\ln(N)}
\end{equation}
and therefore
\begin{equation}
    \alpha^{\mathrm{G}}_{2f} \sim \big[N\ln(N)\big]^{-1},
\end{equation}
where $\alpha^{\mathrm{G}}_{2f}$ denotes $2f$ effective attenuation under Gaussian-distributed $a_n$.

Now suppose the elements of $\boldsymbol{a}$ are uniformly distributed: $a_n \overset{\mathrm{i.i.d.}}{\sim} \mathcal{U}(c, d)$. Then 
\begin{equation}
    \mathrm{P}(|a| > a_{\mathrm{max}}) = \frac{d - a_{\mathrm{max}}}{d - c}\,, \quad c \leq a_{\mathrm{max}} \leq d
\end{equation}
and 
\begin{equation}
    a_{\mathrm{max}} = d - \left(\frac{d - c}{N}\right),
\end{equation}
yielding
\begin{equation}
    \alpha^{\mathrm{U}}_{2f} \sim N^{-1},
\end{equation}
where $\alpha^{\mathrm{U}}_{2f}$ denotes $2f$ effective attenuation under uniformly-distributed $a_n$.

\subsubsection{4f Effective Attenuation Scalings}
In Eq.~(\ref{eq:optical_energy_atten_4f}), if we treat the $\sinc$ factor as roughly constant over $m$ (which is true if we select $\Omega_s$ small enough), then $\alpha_{2f} \propto \left[  \max_m \left |\sum_{n=1}^N a_n \exp{(i2\pi nms \Omega_s)} \right |    \right]^{-2} = \left[  \max_m\left (  \left |\mathcal{A}_m \right | \right )   \right]^{-2}$. The quantity $\mathcal{A}_m$ is the discrete-space Fourier transform of $\boldsymbol{a}$ evaluated at spatial frequencies $\nu_m = m s \Omega_s$.

Suppose the elements of $\boldsymbol{a}$ are Gaussian distributed: $a_n \overset{\mathrm{i.i.d.}}{\sim} \mathcal{N}(0, \sigma^2)$. Then the vector elements $\mathcal{A}_m$ are also Gaussian distributed. Therefore, our previous analysis applies, so that 
\begin{equation}
    \alpha^{\mathrm{G}}_{4f} \sim \big[N\ln(N)\big]^{-1}, 
\end{equation}
where $\alpha^{\mathrm{G}}_{4f}$ denotes $4f$ effective attenuation under Gaussian-distributed $a_n$.

Suppose the elements of $\boldsymbol{a}$ are uniformly distributed: $a_n \overset{\mathrm{i.i.d.}}{\sim} \mathcal{U}(c, d)$. Then for large enough mean $\mu = (c+d)/2$, the zeroth-order Fourier component $\mathcal{A}_0$, which scales as $\mathcal{A}_0 \sim \mu N$, will dominate as the largest magnitude Fourier component. This implies 
\begin{equation}
    \alpha^{\mathrm{U}}_{4f} \sim N^{-2},
\end{equation}
where $\alpha^{\mathrm{U}}_{4f}$ denotes $4f$ effective attenuation under uniformly-distributed $a_n$.

\section{Results}\label{sec:results}

\subsection{Free-Space 2f and 4f Architectures Scale Better to Large MVMs than Integrated UMIs}

\begin{figure*}[htbp]
  \noindent
  \makebox[\textwidth]{\includegraphics[width=1.1\textwidth]{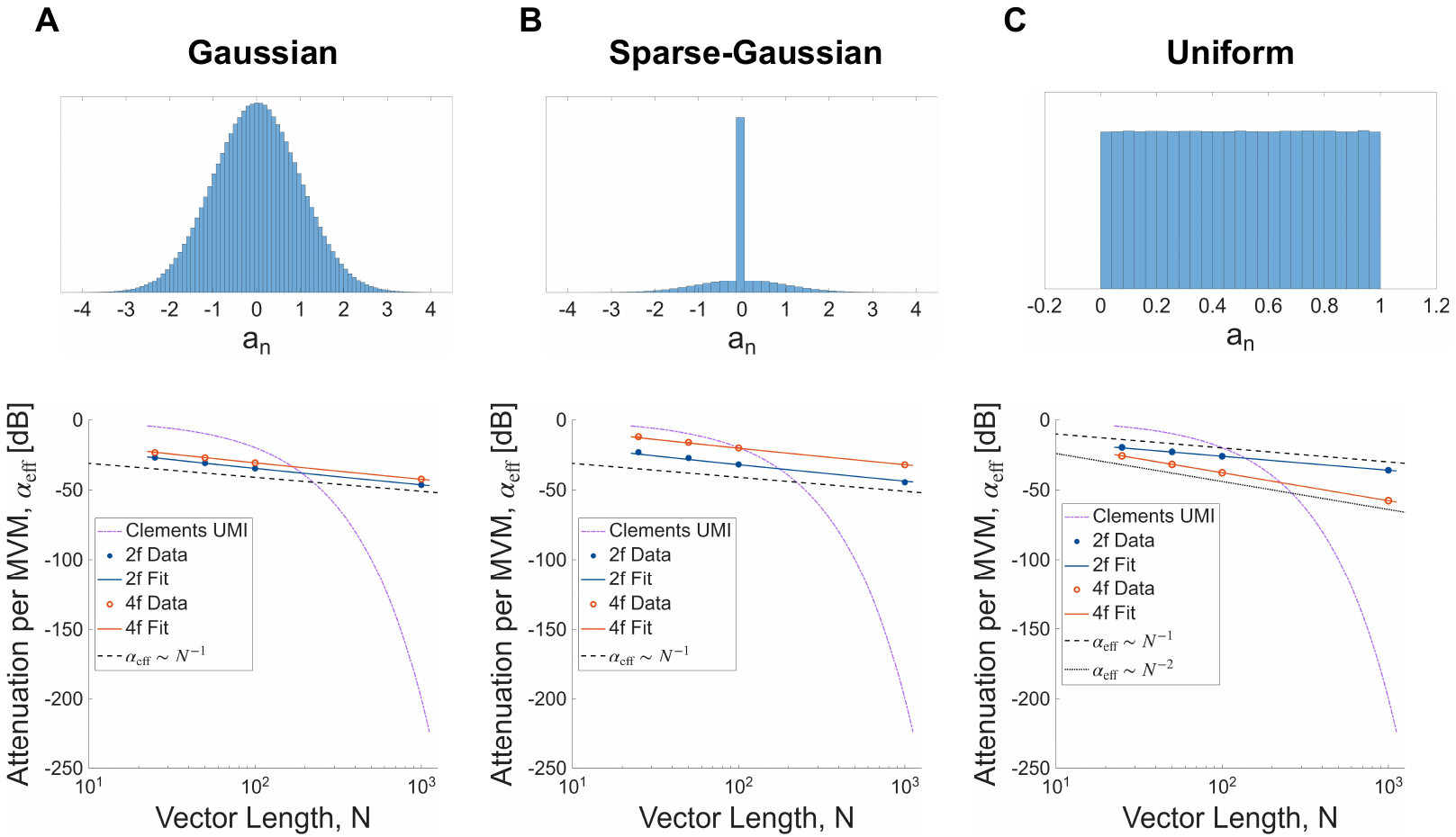}}
  \caption{\label{fig:atten_vs_N} Top: Histograms show the three different distributions of the vector elements $a_n$ described in Sec.~\ref{sec:procedure}. Bottom: Plots of $2f$ and $4f$ effective attenuation per MVM $\alpha_{\mathrm{eff}}$ vs. vector length $N$ of vectors $\boldsymbol{a}$ and $\boldsymbol{v}$ being dotted. $2f$ and $4f$ data are plotted as blue solid and red hollow circles, respectively. Plots correspond to the above distributions of $a_n$. $2f$ data is taken at a slit bandwidth of $\Delta \nu_x = 0.2 \, \mathrm{cm}^{-1}$. Effective attenuation is plotted in $\mathrm{dB}$ and the x-axis uses a log scale. Data markers at each $N$ reflect averages of $\alpha_{\mathrm{eff}}$ over 1000 trials of dot products. Error bars are included but are smaller than the marker size. Solid line least squares fits to the predicted scalings of Sec.~\ref{subsec:scalings_from_stats} are shown, with fit functions and parameters listed in Table \ref{tab:2f_4f_fitting_params}. Comparison is made to the predicted scaling $\alpha_{\mathrm{UMI}}(N) = 10^{-0.02 (N-1)}$ of the Clements UMI performing $N \times N$ MVM with $0.2 \; \mathrm{dB}$ assumed insertion loss per beamsplitter, shown as a purple dash-dotted line. Dashed and dotted black lines provide references to the scalings $\alpha_{\mathrm{eff}} \sim N^{-1}$ and $\alpha_{\mathrm{eff}} \sim N^{-2}$, respectively. \\
  (A) Plots corresponding to Gaussian-distributed $a_n\overset{\mathrm{i.i.d.}}{\sim}\mathcal{N}(0, 1)$.\\
  (B) Plots corresponding to sparse-Gaussian $a_n$ with $90\%$ sparsity.\\
  (C) Plots corresponding to uniform-distributed $a_n \overset{\mathrm{i.i.d.}}{\sim} \mathcal{U}(0, 1)$.
  }
\end{figure*}

\begin{table}[htbp]
\caption{Fitting Functions and Corresponding Parameter Values for $2f$ and $4f$ Data in Fig.~\ref{fig:atten_vs_N}.}
\label{tab:2f_4f_fitting_params}
\centering
\resizebox{\linewidth}{!}{
\begin{tabular}{c c c c}
\hline
 & Gaussian $a_n$ & Sparse-Gaussian $a_n$ & Uniform $a_n$
 \\
\hline
\rule{0pt}{12pt} 
$2f$ Fit & $\alpha_{2f}(N) = b_1\left[N \ln{(N)}\right]^{-1}$  &  $\alpha_{2f}(N) = b_1\left[N \ln{(N)}\right]^{-1}$ & $\alpha_{2f}(N) = b_1 N^{-1}$ \\
$2f$ Fit Parameters & $b_1 = 0.1588 \pm 0.0022$ & $b_1 = 0.29 \pm 0.03$ & $b_1 = 0.2526 \pm 0.0006$\\
\\
$4f$ Fit & $\alpha_{4f}(N) = c_1\left[N \ln{(N)}\right]^{-1}$ & $\alpha_{4f}(N) = c_1\left[N \ln{(N)}\right]^{-1}$ & $\alpha_{4f}(N) = c_1 N ^{-2}$\\
$4f$ Fit Parameters & $c_1 = 0.397 \pm 0.006$ & $c_1 = 4.39 \pm 0.13$ & $c_1 = 1.6857 \pm 0.0015$\\

\hline
\end{tabular}
}
\end{table}

We claim that both $2f$ and $4f$ free-space architectures scale better to large MVM problem sizes (large matrices and vectors) than an integrated photonics UMI mesh. For the the $2f$ and $4f$ architectures, we examine how signal attenuation per $N \times N$ MVM scales with vector length (matrix dimension) $N$ when vector elements $a_n$ are sampled from various statistical distributions. In Fig.~\ref{fig:atten_vs_N} we plot the $2f$ and $4f$ effective attenuation per MVM, $\alpha_{2f}$ and $\alpha_{4f}$, vs. vector length $N$. For each vector length $N \in \{25, 50, 100, 1000\}$, attenuation data are averaged over 1000 trials to obtain each circular data point. $2f$ data is taken at a slit bandwidth of $\Delta \nu_x = 0.2 \, \mathrm{cm}^{-1}$. Effective attenuation per MVM is plotted for the Gaussian, sparse-Gaussian, and uniform distributions of $a_n$ described in Sec.~\ref{sec:procedure}.  Least squares fits matching the predicted scalings of Sec.~\ref{subsec:scalings_from_stats} are provided for 
the $2f$ and $4f$ data of each plot. The fitting functions and corresponding least square parameter values are shown in Table~\ref{tab:2f_4f_fitting_params}.
We see that $2f$ and $4f$ attenuation scales differently with vector length $N$ depending on the statistics of the vector elements $a_n$, verifying the predicted attenuation scalings of Sec.~\ref{sec:pred_scalings}. In particular, $2f$ outperforms $4f$ MVM in the case that $a_n$ is distributed with significant non-zero mean, as in the uniformly-distributed case. Therefore, whether $2f$ or $4f$ experiences more attenuation and is more scalable to increasing $N$ (larger MVM) depends on the statistics of the matrix $\boldsymbol{A}$ being processed.

The $2f$ and $4f$ effective attenuation per $N \times N$ MVM are compared to the predicted attenuation of a Clements universal multiport interferometer (UMI) \cite{Clements:16} implementing $N \times N$ MVM with $0.2 \; \mathrm{dB}$ assumed insertion loss per beamsplitter. The expected attenuation $\alpha_{\mathrm{UMI}}(N) = 10^{-0.02(N-1)}$ is plotted as a dash-dotted purple line. The signal attenuation associated with $2f$ and $4f$ MVM grows much slower with increasing $N$ than for the UMI. Therefore, the $2f$ and $4f$ architectures are more scalable to large MVMs than integrated UMI approaches. This is especially apparent when the matrix is hundreds to thousands of elements wide.

\subsection{4f Error Can Be Reduced by Increasing Modulator Space-Bandwidth Product, but 2f Error Cannot}

\begin{figure*}[h]
  \noindent
  \makebox[\textwidth]{\includegraphics[width=1.1\textwidth]{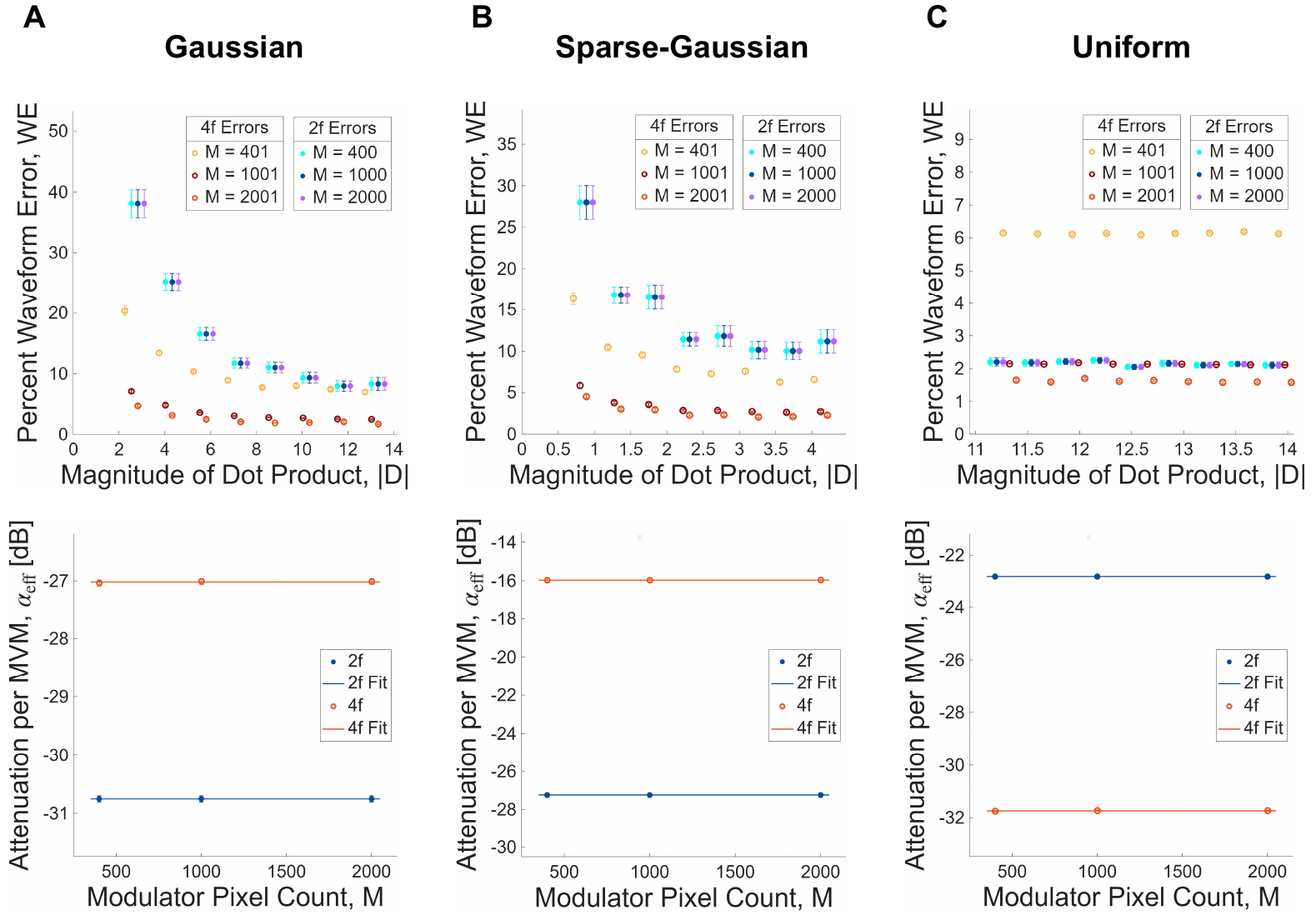}}
  \caption{\label{fig:err_atten_vs_M}
  Top: Plots of $2f$ and $4f$ percent waveform error $\mathrm{WE}$ vs. magnitude of dot product $|D| = |\boldsymbol{a} \cdot \boldsymbol{v}|$ for varying modulator pixel count $M$ at vector length $N = 50$ for each distribution of $a_n$. $2f$ and $4f$ data are plotted as cool-colored solid and hot-colored hollow circles, respectively. $2f$ error data is taken at a slit bandwidth of $\Delta \nu_x = 0.2 \, \mathrm{cm}^{-1}$. Waveform error data are binned according to dot product magnitude and averaged for each bin to obtain the data points and error bars. Some error bars are smaller than the marker size. The ranges of dot product magnitudes shown in each plot enclose approximately $95\%$ of the data taken for each distribution. Bottom: Corresponding plots of $2f$ and $4f$ effective attenuation per MVM $\alpha_{\mathrm{eff}}$ vs. modulator pixel count $M$. Effective attenuation is plotted in $\mathrm{dB}$. $2f$ attenuation data is taken at a slit bandwidth of $\Delta \nu_x = 0.2 \, \mathrm{cm}^{-1}$. Data markers at each $M$ are averages, with error bars included. Solid lines of zero slope are fitted to the data.\\
  (A) Plots corresponding to Gaussian-distributed $a_n\overset{\mathrm{i.i.d.}}{\sim}\mathcal{N}(0, 1)$. \\
  (B) Plots corresponding to sparse-Gaussian $a_n$ with $90\%$ sparsity.\\
  (C) Plots corresponding to uniform-distributed $a_n \overset{\mathrm{i.i.d.}}{\sim} \mathcal{U}(0, 1)$.}
\end{figure*}

We claim that computational error in the $4f$ architecture tends to decrease to lower levels when increasing the number of modulator pixels (i.e. increasing modulator space-bandwidth product), whereas $2f$ errors are insensitive. We also claim that optical signal attenuation is unaffected by modulator pixel count. We fix vector length to $N = 50$ and measure how error and signal attenuation are affected by increasing modulator pixel count $M$. In the top plots of Fig.~\ref{fig:err_atten_vs_M} we compare $2f$ and $4f$ percent waveform error $\mathrm{WE}$ (defined in Sec.~\ref{sec:definitions}) vs. magnitude of the optically computed dot product $|D| = |\boldsymbol{a} \cdot \boldsymbol{v}|$ for modulator pixel counts $M_{2f} \in \{400, 1000, 2000\}$ and $M_{4f} \in \{401, 1001, 2001\}$. $2f$ data is taken at a slit bandwidth of $\Delta \nu_x = 0.2 \, \mathrm{cm}^{-1}$. Percent waveform error varies with dot product magnitude since deviations from the desired target output field will more greatly affect the "solution" encoded in an optical signal of lower intensity (smaller dot product) than one of higher intensity. We see that, across the different distributions of $a_n$, $4f$ waveform error is reduced by increasing $M$ while $2f$ is not affected. Increasing $M$ provides diminishing improvements to $4f$ error since the modulator pixels encode the Fourier transform of the vector $\boldsymbol{a}$ (see Sec.~\ref{subsec:4f_MVM_analytic}). Adding more pixels means adding higher-frequency samples of the Fourier transform, which tend to provide smaller and smaller corrections to the approximate encoding of $\boldsymbol{a}$. $2f$ error is unaffected by $M$ since adding more modulator pixels does not improve the accuracy of our encoding of $\boldsymbol{a}$ using a superpixel encoding approach (see Sec.~\ref{subsec:2f_MVM_analytic}). 

In the bottom plots of Fig.~\ref{fig:err_atten_vs_M} we show $2f$ and $4f$ average effective attenuation per MVM $\alpha_{\mathrm{eff}}$ vs. modulator pixel count $M$. The data in the bottom plots are plotted at the same $M$ values as the $2f$ and $4f$ data of the top plots. Neither $2f$ nor $4f$ attenuation per MVM is noticeably affected by increasing modulator pixel count $M$. In the $2f$ case, this is true because the input optical signal encoding $\boldsymbol{v}$ is shaped to perfectly match the pixel pattern on the modulator (see Sec.~\ref{subsec:2f_MVM_analytic}), so no optical energy is lost to the modulator regardless of pixel count. In the $4f$ case, for large enough pixel counts $M$, almost all of the incoming signal energy falls within the modulator width. Adding modulator pixels provides smaller and smaller corrections to the signal energy passed by the modulator, hence a very weak dependence of attenuation on $M$.    

Combining these observations, $4f$ error can be reduced by increasing modulator space-bandwidth product (via pixel count $M$) while keeping attenuation per MVM practically constant. On the other hand, $2f$ error cannot be reduced by increasing modulator space-bandwidth product.

\subsection{Lower 2f Error Must Be Traded for Higher Attenuation}

\begin{figure}[htbp]
  \noindent
  \makebox[\textwidth]{\includegraphics[width=1.1\textwidth]{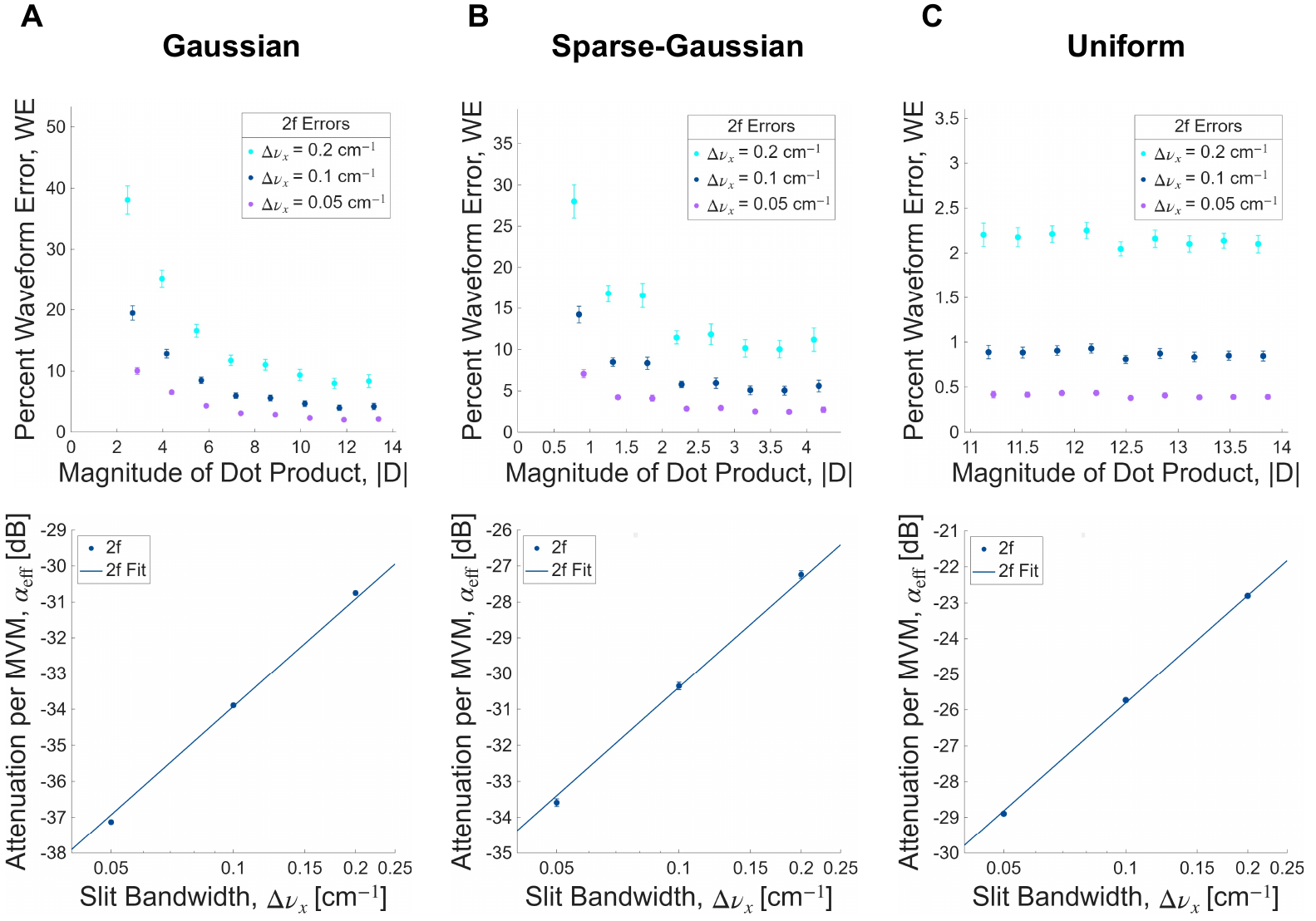}}
  \caption{\label{fig:err_atten_vs_SW}
  Top: Plots of $2f$ percent waveform error $\mathrm{WE}$ vs. magnitude of dot product $|D| = |\boldsymbol{a} \cdot \boldsymbol{v}|$ for varying slit bandwidth $\Delta \nu_x$ at vector length $N = 50$ for each distribution of $a_n$. $2f$ error data is taken at $M = 2000$ modulator pixels. Waveform error data are binned according to dot product magnitude and averaged for each bin to obtain the data points (with error bars). Some error bars are smaller than the marker size. The ranges of dot product magnitudes shown in each plot enclose approximately $95\%$ of the data taken for each distribution. Bottom: Corresponding plots of $2f$ effective attenuation per MVM $\alpha_{\mathrm{eff}}$ vs. slit bandwidth $\Delta \nu_x$. Effective attenuation is plotted in $\mathrm{dB}$. Data markers at each $\Delta \nu_x$ are averages, with error bars included. Least squares fits are applied to the data for fitting function $\alpha_{2f}(\Delta \nu_x) = b_1 \Delta \nu_x$.\\
  (A) Plots corresponding to Gaussian-distributed $a_n\overset{\mathrm{i.i.d.}}{\sim}\mathcal{N}(0, 1)$.\\
  (B) Plots corresponding to sparse-Gaussian $a_n$ with $90\%$ sparsity.\\
  (C) Plots corresponding to uniform-distributed $a_n \overset{\mathrm{i.i.d.}}{\sim} \mathcal{U}(0, 1)$.
  }
\end{figure}

We claim that computational error can be reduced in the $2f$ architecture by decreasing the slit bandwidth, but that this also leads to more signal attenuation. In the top plots of Fig.~\ref{fig:err_atten_vs_SW} we show $2f$ percent waveform error $\mathrm{WE}$ vs. magnitude of the optically computed dot product $|D| = |\boldsymbol{a} \cdot \boldsymbol{v}|$ for slit bandwidths $\Delta \nu_x \in \{0.05 \ \mathrm{cm}^{-1}, 0.1 \,  \mathrm{cm}^{-1}, 0.2 \,  \mathrm{cm}^{-1}\}$. $2f$ data is taken at modulator pixel count $M = 2000$. Percent waveform error decreases as the slit bandwidth is decreased. This is because, as the slit bandwidth decreases, we select out a sliver of the $2f$ field that is closer and closer to the desired zeroth-order Fourier component encoding the dot product solution.

In the bottom plots of Fig.~\ref{fig:err_atten_vs_SW} we show the average effective attenuation per MVM $\alpha_{\mathrm{eff}}$ corresponding to the top plots. Decreasing slit bandwidth lets less light exit the $2f$ MVM system, thus linearly reducing the $\alpha_{\mathrm{eff}}$ factor. Least squares fits to the function $\alpha_{2f}(\Delta \nu_x) = b_1 \Delta \nu_x$ are shown, with $b_1 = (4.05 \pm 0.10) \times 10^{-3}$, $b_1 = (9.14 \pm 0.20) \times 10^{-3}$, and $b_1 = (2.63 \pm 0.03) \times 10^{-2}$ for the Gaussian, sparse-Gaussian, and uniform cases, respectively. We see that to achieve lower $2f$ error, we must decrease the slit bandwidth and consequently sacrifice for more attenuation per MVM.

\section{Discussion}

Our results suggest that free-space $2f$ and $4f$ MVM systems are more scalable to large matrix sizes than current integrated photonic UMI approaches. This is because, for $N \times N$ MVM, the Clements \cite{Clements:16} UMI spreads $N^2$ matrix weights across a multilayer mesh of $N(N-1)/2$ beamsplitters. Each optical mode must cross approximately $N-1$ layers of beamsplitters before exiting the mesh, accumulating insertion loss for each beamsplitter. Thus, the effective attenuation $\alpha_{\mathrm{eff}}$ experienced by the signal grows exponentially with matrix dimension $N$. This $O(N)$ scaling of the number of modulating layers for MVM was also remarked as a reason for the limited scalability of $2$D integrated approaches for optical neural networks in \cite{bernstein2023single}. We expect deep diffractive networks for MVM \cite{arb_linear_trans} to exhibit similar exponential scaling due to losses at each diffractive interface. In contrast to UMIs, the $2f$ and $4f$ architectures require only a single layer of modulation and mixing for all matrix sizes, and thus do not accumulate loss exponentially with $N$. For $2f$ and $4f$ $N \times N$ MVM, copies of the $N$ input modes, corresponding to the elements of input vector $\boldsymbol{v}$, are made for each row of the matrix. All copies are modulated by the matrix weights in a single modulating layer, after which a lens mixes the modulated modes by transforming to the spatial frequency (Fourier) basis. After modulating and mixing, we use a slit in the output plane to select out the output modes encoding the desired MVM solution. With the $2f$ and $4f$ architectures, we decrease the depth of the MVM network by modulating in a single layer and throwing away undesired optical modes at the output, in contrast to the UMI and diffractive network systems which do not intentionally filter out modes. We can roughly explain the attenuation scaling of $2f$ and $4f$ architectures using a naive line of reasoning: if we have $N$ input modes per matrix row and select only one desired output mode per matrix row, we would expect a $1/N$ attenuation scaling. This partially explains the scalings we see in Fig.~\ref{fig:atten_vs_N}, where deviations from $1/N$ scaling can be attributed to the rescaling of the modulator transmittance (see Sec.~\ref{subsec:modeling_the_modulator}). 

We observe from our results that, when optically computing $\boldsymbol{A}\boldsymbol{v}$, $2f$ and $4f$ attenuation scaling with $N$ depends on the statistics of the row vectors of matrix $\boldsymbol{A}$. As a consequence, we cannot make a general suggestion of whether $2f$ or $4f$ scales more favorably with matrix dimension $N$. In Sec.~\ref{sec:pred_scalings}, we derived expressions for $2f$ and $4f$ effective attenuation (see Eq.~(\ref{eq:optical_energy_atten_2f}) and Eq.~(\ref{eq:optical_energy_atten_4f})) that show attenuation depends on "modulator transmittance rescaling", described in the following. We assumed the modulator had a maximum amplitude gain of $1$ and used the full dynamic range of the modulator. A row of the modulator transmittance mask encodes a vector $\boldsymbol{a}$ in the $2f$ case or its Fourier transform $\boldsymbol{\mathcal{A}} = \mathcal{F}\{\boldsymbol{a}\}$ in the $4f$ case. Since the maximum transmittance value is constrained by the maximum amplitude gain of the modulator, we must rescale $\boldsymbol{a}$ or $\boldsymbol{\mathcal{A}}$ (namely by $1/\max(\boldsymbol{a})$ or $1/\max(\boldsymbol{\mathcal{A}})$) so the modulator can encode them in its transmittance (see Sec.~\ref{subsec:modeling_the_modulator}). Since these $2f$ and $4f$ transmittance rescaling factors will depend on the statistics of the elements of $\boldsymbol{a}$, we see differences between $2f$ and $4f$ attenuation scaling for different distributions of $a_n$ (see Sec.~\ref{subsec:scalings_from_stats}). As Fig.~\ref{fig:atten_vs_N}(C) suggests, $2f$ MVM has more favorable scaling for $a_n$ distributed with significant non-zero mean $\mu$ because, while $\max_{n}(a_n)$ may not grow significantly with $N$, $\max_{n}(\mathcal{A}_n)$ will grow as $\mu N$. This means the modulator rescaling factor of $4f$ will outgrow that of $2f$, leading to more aggressive attenuation scaling for $4f$. 

Our results furthermore demonstrate that $4f$ error can be reduced by increasing modulator space-bandwidth product without affecting attenuation, while lower $2f$ error must be traded for larger attenuation. Increasing the modulator space-bandwidth product by increasing the number of modulator pixels improves the sampling of $\boldsymbol{\mathcal{A}} = \mathcal{F}\{\boldsymbol{a}\}$ in the $4f$ setup. Better sampling in the Fourier plane lowers computation error while not increasing signal loss, since more spatial frequencies are passed through the modulator. On the other hand, the way we encoded the $2f$ transmittance and input field using superpixels means increasing the modulator's space-bandwidth product does not improve our encoding of $\boldsymbol{a}$ and so does not reduce computation error. To reduce $2f$ error, the slit bandwidth must be narrowed to select spatial frequencies closer to the desired solution at $\nu_x = 0$. Consequently, light at other frequencies is lost, leading to more signal loss per MVM.   

Taken together, our results suggest that free-space $2f$ and $4f$ architectures are more scalable than UMIs to large MVM problem sizes. This is important because photonic hardware will likely need to process large matrices to gain advantage over electronics, for which latency grows with data size due to the von Neumann bottleneck. The improved attenuation scaling of $2f$ and $4f$ MVM also allow for cascaded computation at large matrix dimensions. Additionally, the results frame the $4f$ architecture as a more flexible architecture than $2f$, being able to reduce error without sacrificing for higher signal attenuation. Overall, we suggest that photonic computing architectures should minimize the number of modulation and mixing layers (as in the $2f$ and $4f$ architectures) to optimize for scalability and cascadability, which can be critical for realizing impactful optical computing hardware.

Next steps include experimentally verifying the scaling and error behaviors of the $2f$ and $4f$ architectures predicted by our simulations. Given the large size of free-space setups, another direction could be implementing the $2f$ and $4f$ architectures using compact nanophotonic or integrated designs. 

\section*{Disclosures} The authors declare no conflicts of interest.

\section*{Data availability} Data underlying the results presented in this paper are not publicly available at this time but may be obtained from the authors upon reasonable request.

\clearpage

\input{supplement}

\bibliography{main_refs}

\end{document}

%% file: supplement.tex
\title{Attenuation scaling and error analysis of 2f and 4f architectures for free-space optical matrix-vector multiplication: supplement}

\medskip

\setcounter{section}{0}
\renewcommand{\thesection}{S\arabic{section}}

\setcounter{equation}{0}
\renewcommand{\theequation}{S\arabic{equation}}

\section{Two Methods for MVM Using Fourier Transforms} \label{sec: Methods 1 and 2}

We wish to perform the matrix-vector multiplication (MVM)
\begin{equation} \label{eq:dot_prod}
    \boldsymbol{A} \boldsymbol{v} = \begin{pmatrix}
        A_{11} & A_{12} & \ldots & A_{1N} \\
        A_{21} & A_{22} & \ldots & A_{2N} \\
        \vdots & \vdots & & \vdots \\
        A_{M1} & A_{M2} & \ldots & A_{MN}
    \end{pmatrix}_{M \times N}  \begin{pmatrix}
        v_1 \\
        v_2 \\
        \vdots \\
        v_N
    \end{pmatrix} = \begin{pmatrix}
        \sum_{j=1}^N A_{1j}v_j \\
        \sum_{j=1}^N A_{2j}v_j \\
        \vdots \\
        \sum_{j=1}^N A_{Mj}v_j
    \end{pmatrix} = \boldsymbol{b},
\end{equation}
where $\boldsymbol{A}$ has dimensions $M \times N$ and $\boldsymbol{v}$ and $\boldsymbol{b}$ are column vectors of dimensions $N$ and $M$, respectively. To perform MVM, column vector $\boldsymbol{v}$ is transposed and multiplied elementwise with each row of matrix $\boldsymbol{A}$. The products of each row are then summed together to give the entries of the resulting vector $\boldsymbol{b}$. Summation along the rows is a many-to-one operation where multiple inputs, the elements of the matrix row, are mapped to a single output, their sum. Many-to-one summation is not possible using lossless, passive linear optics, so alternative approaches must be used to achieve summation in optical MVM. Therefore, we describe two methods for achieving MVM (i.e. for producing the same elements as in $\boldsymbol{b}$) without using many-to-one summation.

\subsection{Method 1} \label{subsec:method1}
Method 1 transposes column vector $\boldsymbol{v}$ and copies it $M$ times to produce an $M \times N$ matrix, $\boldsymbol{V}$, matching the dimensions of $\boldsymbol{A}$:
 \begin{equation} \label{eq:V_matrix}
    \boldsymbol{V} = \begin{pmatrix}
        v_1 & v_2 & \ldots & v_N \\
        v_1 & v_2 & \ldots & v_N \\
        \vdots & \vdots & & \vdots \\
        v_1 & v_2 & \ldots & v_N
    \end{pmatrix}_{M \times N}.
\end{equation}
Then, an elementwise product between $\boldsymbol{A}$ and $\boldsymbol{V}$ is taken
\begin{equation} \label{eq:elementwise_prod}
    \boldsymbol{C} = \boldsymbol{A} \odot \boldsymbol{V} =  
    \begin{pmatrix}
        A_{11}v_1 & A_{12}v_2 & \ldots & A_{1N}v_N \\
        A_{21}v_1 & A_{22}v_2 & \ldots & A_{2N}v_N \\
        \vdots & \vdots & & \vdots \\
        A_{M1}v_1 & A_{M2}v_2 & \ldots & A_{MN}v_N \\
    \end{pmatrix}_{M \times N}, \\ 
\end{equation}
followed by an $N$-point discrete Fourier transform (DFT) along the rows of this matrix to obtain
\begin{equation} \label{eq:FT_of_elementwise_prod}
    \mathcal{F}_{N,\mathrm{row}} \{ \boldsymbol{C} \}  =  \scalebox{0.8}{$
    \begin{pmatrix}
        \sum_{j=1}^{N} C_{1j}   & \sum_{j=1}^{N} C_{1j} \exp\!{ \left [-i2\pi\frac{(j-1)}{N} \right]} & \ldots & \sum_{j=1}^{N} C_{1j} \exp\!{\left [-i2\pi\frac{(N-1) (j-1)}{N}\right ]} \\
        \sum_{j=1}^{N} C_{2j}   & \sum_{j=1}^{N} C_{2j} \exp\!{\left [-i2\pi\frac{(j-1)}{N} \right ]} & \ldots & \sum_{j=1}^{N} C_{2j} \exp\!{\left [-i2\pi\frac{(N-1) (j-1)}{N} \right ]} \\
        \vdots & \vdots & & \vdots \\
        \sum_{j=1}^{N} C_{Mj}   & \sum_{j=1}^{N} C_{Mj}  \exp\!{\left [-i2\pi\frac{(j-1)}{N}\right ]} & \ldots & \sum_{j=1}^{N} C_{Mj} \exp\!{\left [-i2\pi\frac{(N-1) (j-1)}{N}\right ]} \\
    \end{pmatrix}_{M \times N} $}.
    \\
\end{equation}
The first column of this matrix, corresponding to the zeroth-order Fourier component of each row of $\boldsymbol{C}$, is the desired MVM solution:
\begin{equation} \label{eq:2f_MVM_solution}
   (\mathcal{F}_{N,\mathrm{row}} \{ \boldsymbol{C} \})_{:,1} = \begin{pmatrix}
        \sum_{j=1}^{N} C_{1j}   \\
        \sum_{j=1}^{N} C_{2j} \\
        \vdots \\
        \sum_{j=1}^{N} C_{Mj}   \\
    \end{pmatrix}  = \begin{pmatrix}
        \sum_{j=1}^N A_{1j}v_j \\
        \sum_{j=1}^N A_{2j}v_j \\
        \vdots \\
        \sum_{j=1}^N A_{Mj}v_j
    \end{pmatrix} = \boldsymbol{b}.
\end{equation}

\subsection{Method 2} \label{subsec:method2}
Method 2 first reverses the order of the elements in all rows of $\boldsymbol{V}$ to give a new matrix $\boldsymbol{V}'$. Then, each row of $\boldsymbol{V}'$ is convolved with each row of $\boldsymbol{A}$, yielding a matrix of dimensions $M \times (2N-1)$:
\begin{equation} \label{eq:method_2_convolution}
    \boldsymbol{A} (*_{\mathrm{row}}) \boldsymbol{V}' = \scalebox{0.70}{$ \begin{pmatrix}
        A_{11}v_N & A_{11}v_{N-1} + A_{12}v_N & \ldots & \sum_{j=1}^N A_{1j}v_j & \ldots &  A_{1,N-1}v_{1} + A_{1N}v_2 & A_{1N}v_1 \\
        A_{21}v_N & A_{21}v_{N-1} + A_{22}v_N & \ldots & \sum_{j=1}^N A_{2j}v_j & \ldots &  A_{2,N-1}v_{1} + A_{2N}v_2 & A_{2N}v_1\\
        \vdots & \vdots & & \vdots & & \vdots & \vdots \\
        A_{M1}v_N & A_{M1}v_{N-1} + A_{M2}v_N & \ldots & \sum_{j=1}^N A_{Mj}v_j & \ldots &  A_{M,N-1}v_{1} + A_{MN}v_2 & A_{MN}v_1 \\
    \end{pmatrix}_{M \times (2N-1)}$},
\end{equation}
where 
\begin{equation} \label{eq:V_reversed}
    \boldsymbol{V}' = \begin{pmatrix}
        v_N & v_{N-1} & \ldots & v_1 \\
        v_N & v_{N-1} & \ldots & v_1 \\
        \vdots & \vdots & & \vdots \\
        v_N & v_{N-1} & \ldots & v_1 \\
    \end{pmatrix}_{M \times N}.
\end{equation}
The middle ($N$th) column of the matrix in Eq.~(\ref{eq:method_2_convolution}) is the desired MVM solution $\boldsymbol{b}$:
\begin{equation} \label{eq:4f_MVM_solution}
   (\boldsymbol{A} *_{\mathrm{row}} \boldsymbol{V}')_{:,N} = \begin{pmatrix}
        \sum_{j=1}^N A_{1j}v_j \\
        \sum_{j=1}^N A_{2j}v_j \\
        \vdots \\
        \sum_{j=1}^N A_{Mj}v_j
    \end{pmatrix} = \boldsymbol{b}.
\end{equation}

\section{Additional Wave Optics Calculations}

\subsection{2f System} \label{sec: 2f Diffraction Calculation}
In this calculation, we show that a $2f$ system using a cylindrical lens performs a Fourier transform along one spatial dimension. If using a spherical lens, the calculation can be generalized to show that a two-dimensional Fourier transform is performed. We refer to the $2f$ schematic of Fig.1(A) in the main text for coordinate axes labels $x_i$ for the planes of interest. Since the cylindrical lens focuses only along the $x$-dimension, we compute diffractive effects and fields only along $x$, holding $y$ constant. 

We wish to calculate the output field $U_{\mathrm{out}}(x_2)$ given the field $U_1(x_1)$ immediately following the modulator plane $x_1y_1$. From Fig.1(A), we see that, after an input scalar field $U_{\mathrm{in}}$ passes through a modulator transmittance mask $t_{2f}$, the field immediately following the modulator plane is 
\begin{equation}
U_1(x_1) = t_{2f}(x_1)\, U_{\mathrm{in}}(x_1).
\end{equation}
The field $U_1(x_1)$ lies in the front focal plane of the cylindrical lens. To calculate the field after propagating a distance $d=f$ to the plane of the cylindrical lens, we multiply the angular spectrum of this field, $\mathcal{F}_{x}\{U_1\}(\nu_x)$, by the one-dimensional form of the transfer function for free-space Fresnel diffraction by a distance $d$ (see Goodman~\cite[p.~72]{goodman1996fourier})
\begin{equation}
    H = \exp\!{(i2\pi d/\lambda)} \, \exp\!{\left(-i\pi \lambda d \nu_x^2\right)}.
\end{equation}
We then obtain the angular spectrum of the field $U_2$ in the plane immediately before the thin lens
\begin{equation}
    \mathcal{F}_{x}\{U_2\}(\nu_x) = \exp\!{(i2\pi d/\lambda)} \, \exp\!{\left(-i\pi \lambda d \nu_x^2\right)} \, \mathcal{F}_{x}\{U_1\}(\nu_x),
\end{equation}
where $\nu_x$ is the spatial frequency for the $x$-dimension. Next, the field passes through the thin cylindrical lens, which applies a quadratic phase mask, and propagates a distance $f$ to the output plane. Following Goodman~\cite[Ch.~5]{goodman1996fourier}, the output field is related to the field immediately before the lens according to 
\begin{equation}
    U_{\mathrm{out}}(x_2) = \frac{\exp\!{(i2\pi f/\lambda})}{i\sqrt{\lambda f}} \exp\!{\left[i\pi x_2^2/(\lambda f)\right]}\,\mathcal{F}_{x}\{U_2\}\!\left[\nu_x = x_2/(\lambda f)\right].
\end{equation} 
Substituting and simplifying, we obtain the output field
\begin{equation}
    \begin{aligned}
        U_{\mathrm{out}}(x_2) & = \frac{\exp\!{(i2\pi f/\lambda})}{i \sqrt{\lambda f}}\exp\!{\left[i\pi x_2^2/(\lambda f)\right]} \, \mathcal{F}_{x}\{U_2\}\!\left[\nu_x = x_2/(\lambda f)\right]\\
        & = \begin{multlined}[t]
        \frac{\exp\!{(i2\pi f/\lambda)}}{i\sqrt{\lambda f}}\exp\!{\left[i\pi x_2^2/(\lambda f)\right]} \exp\!{(i2\pi d/\lambda)} \exp\!{\left[-i\pi d x_2^2 / (\lambda f^2)\right]} \\ \times \mathcal{F}_{x}\{U_1\}\!\left[\nu_x = x_2/(\lambda f)\right] 
        \end{multlined}
        \\
        & = \begin{multlined}[t]
        \frac{\exp\!{[i2\pi (f + d)/\lambda]}}{i\sqrt{\lambda f}}\exp\!{\left[i\pi \! \left(1 - \frac{d}{f}\right)x_2^2/(\lambda f)\right]} \\
        \times \int_{-\infty}^{\infty} U_1(\xi) \exp\!{\left[-i2\pi x_2 \xi/(\lambda f)\right]} \, \mathrm{d}\xi 
        \end{multlined}
        \\
        & = \begin{multlined}[t]
        \frac{\exp\!{[i2\pi (f + d)/\lambda]}}{i\sqrt{\lambda f}} \exp\!{\left[i\pi \!\left(1 - \frac{d}{f}\right)x_2^2/(\lambda f)\right]}  \\ \times \int_{-\infty}^{\infty} t_{2f}(\xi) \, U_{\mathrm{in}}(\xi) \exp\!{\left[-i2\pi x_2 \xi/(\lambda f)\right]} \, \mathrm{d}\xi.
        \end{multlined}
    \end{aligned}
\end{equation}
We see that, for $d=f$ in our configuration, the output field $U_{\mathrm{out}}$ is the Fourier transform of the field immediately following the modulator plane $U_1$, up to a complex constant factor.

\subsection{4f System} \label{sec: 4f Diffraction Calculation}
In the following, we demonstrate that a $4f$ system yields an output field which is a convolution between an input optical field $U_{\mathrm{in}}$ and the Fourier transform of the modulator transmittance mask between the two cylindrical lenses. We wish to calculate the output field $U_{\mathrm{out}}(x_3)$, considering how the field evolves along only one spatial dimension $x$, since we use two cylindrical lenses that focus only along the $x$-dimension. We refer to Fig.1(B) in the main text for coordinate axes labels, but modify the setup so that the first cylindrical lens has focal length $f_1$ and the second has $f_2$. 

The input optical field $U_{\mathrm{in}}$ propagates a distance $f_1$ to the first cylindrical lens, passes through the lens, and propagates a distance $f_1$ to the modulator plane $x_2y_2$. We borrow the results of Eq.~(S5), finding the scalar field in the plane immediately before the modulator to be 
\begin{equation}
    \begin{aligned}
    U_1(x_2) & = \frac{1}{\sqrt{\lambda f_1}} \mathcal{F}_x\{ U_{\mathrm{in}} \}\!\left [\nu_x = x_2/(\lambda f_1)\right] \\ 
    & = \frac{1}{\sqrt{\lambda f_1}} \int_{-\infty}^{\infty} U_{\mathrm{in}}(x_1) \exp\!{\left[-i2\pi x_2 x_1/(\lambda f_1)\right]} \, \mathrm{d}x_1,
    \end{aligned}
\end{equation}
where we have dropped all constant-phase exponential factors. The field immediately following the modulator plane is 
\begin{equation}
    U_2(x_2) = t_{4f}(x_2) \,  U_1(x_2) = \frac{1}{\sqrt{\lambda f_1}} \, t_{4f}(x_2) \, \mathcal{F}_x\{ U_{\mathrm{in}} \} \!\left[\nu_x = x_2/(\lambda f)\right].
\end{equation} The output field is then
\begin{equation}
    \begin{aligned}
        U_{\mathrm{out}}(x_3) &= \frac{1}{\lambda \sqrt{f_1 f_2}} \int_{-\infty}^{\infty} t_{4f}(x_2) \, \mathcal{F}_x\{ U_{\mathrm{in}} \}\left[x_2/(\lambda f_1) \right] \, \exp\!{\left[-i2\pi x_3 x_2/(\lambda f_2)\right]} \, \mathrm{d}x_2 \\
        & = \begin{multlined}[t] 
        \frac{1}{\lambda \sqrt{f_1 f_2}} \int_{-\infty}^{\infty} t_{4f}(x_2) \left( \int_{-\infty}^{\infty} U_{\mathrm{in}}(x_1) \exp\!{\left[-i2\pi x_1 x_2/(\lambda f_1)\right]} \, \mathrm{d}x_1 \right) \\ \times \exp\!{\left[-i2\pi x_3 x_2/(\lambda f_2)\right]} \, \mathrm{d}x_2 
        \end{multlined} \\
        & = \begin{multlined}[t]
        \sqrt{\frac{f_1}{f_2}} \int_{-\infty}^{\infty} t_{4f}(\lambda f_1 \nu) \left( \int_{-\infty}^{\infty} U_{\mathrm{in}}(x_1) \exp\!{(-i2\pi \nu x_1)} \, \mathrm{d}x_1 \right) \\ \times \exp\!{\left(-i2\pi \frac{f_1 x_3 \nu}{f_2}\right)} \, \mathrm{d}\nu; \quad \nu = \frac{x_2}{\lambda f_1} 
        \end{multlined} \\
        & = \sqrt{\frac{f_1}{f_2}} \int_{-\infty}^{\infty} U_{\mathrm{in}}(x_1)  \left( \int_{-\infty}^{\infty} t_{4f}(\lambda f_1 \nu) \exp\!{\left[-i2\pi\left(x_1 +  \frac{f_1x_3}{f_2}\right) \nu\right]} \, \mathrm{d}\nu \right) \, \mathrm{d}x_1 \\
        & = \sqrt{\frac{f_1}{f_2}} \int_{-\infty}^{\infty} U_{\mathrm{in}}(x_1) \, \mathcal{F}_\nu \{ T \}\!\left ( x_1 +  \frac{f_1x_3}{f_2} \right ) \, \mathrm{d}x_1; \quad T(\nu) = t_{4f}(\lambda f_1 \nu) \\
        & = \sqrt{\frac{f_1}{f_2}} \int_{-\infty}^{\infty} U_{\mathrm{in}}(-x_1) \, \mathcal{F}_\nu \{ T \} \!\left(\frac{f_1x_3}{f_2} - x_1 \right ) dx_1\\
        & = \sqrt{\frac{f_1}{f_2}} \big[ U_{\mathrm{in}}(-x_1) \ast \mathcal{F}_\nu\{ T\}(x_1) \big]\!\left(\frac{f_1 x_3}{f_2}\right), 
    \end{aligned}  
\end{equation}
which is a scaled convolution between the input scalar field and the Fourier transform of the modulator transmittance function.